\documentclass[twocolumn]{aastex63}

\usepackage{graphicx}	
\usepackage{color}
\usepackage{amsmath}	% Advanced maths commands
\usepackage{amssymb}	% Extra maths symbols
\usepackage{mathrsfs}
\usepackage{mathrsfs}
\usepackage{multirow}

\newcommand{\msun}{M_\odot}

\newcommand{\lsun}{L_\odot}

\newcommand{\cc}{{\rm cm}^{-3}}

\newcommand{\mum}{\mu {\rm m}}
\newcommand{\kms}{{\rm km~s}^{-1}}

\newcommand{\K}{{\rm K}}
\newcommand{\beq}{\begin{equation}}
\newcommand{\eeq}{\end{equation}}

\shorttitle{LRDs from Binary BHs}
\shortauthors{Inayoshi et al.}

\begin{document}

\title{The Emergence of Little Red Dots from Binary Massive Black Holes}

\correspondingauthor{Kohei Inayoshi}
\email{inayoshi@pku.edu.cn}

\author[0000-0001-9840-4959]{Kohei Inayoshi}
\affiliation{Kavli Institute for Astronomy and Astrophysics, Peking University, Beijing 100871, China}

\author[0000-0002-4569-9009]{Jinyi Shangguan}
\affiliation{Kavli Institute for Astronomy and Astrophysics, Peking University, Beijing 100871, China}

\author[0000-0002-4569-9009]{Xian Chen}
\affiliation{Department of Astronomy, School of Physics, Peking University, Beijing 100871, China}
\affiliation{Kavli Institute for Astronomy and Astrophysics, Peking University, Beijing 100871, China}

\author[0000-0001-9840-4959]{Luis C. Ho}
\affiliation{Kavli Institute for Astronomy and Astrophysics, Peking University, Beijing 100871, China}
\affiliation{Department of Astronomy, School of Physics, Peking University, Beijing 100871, China}

\author[0000-0003-3633-5403]{Zolt\'an Haiman}
\affiliation{Department of Astronomy, Columbia University, MC 5246, 538 West 120th Street, New York, NY 10027, USA}
\affiliation{Department of Physics, Columbia University, MC 5255, 538 West 120th Street, New York, NY 10027, USA}
\affiliation{Institute of Science and Technology Austria, AM Campus 1, Klosterneuburg 3400, Austria}

\begin{abstract}
Little red dots (LRDs) are a newly identified class of broad-line active galactic nuclei (AGN)
with a distinctive v-shape spectrum characterized by red optical and blue UV continuum emission.
Their high abundance at redshifts of $z\sim6-8$ and decline at lower redshifts suggest a transient origin.
We propose that the spectral shape of LRDs originates from compact binary black hole systems, 
where each black hole is surrounded by a mini-disk and embedded in a larger circum-binary disk. 
With a binary separation of $\lesssim 10^3$ Schwarzschild radii, the Wien tail of a $T\simeq 5000~\K$ blackbody spectrum
at the inner edge of the circum-binary disk produces the red optical emission, while the mini-disks power the UV continuum. 
Binary torques carve out a gap between the circum-binary disk and mini-disks, setting the turnover wavelength 
of the v-shaped spectrum around the Balmer limit.
This scenario naturally reproduces LRD spectra requiring only modest dust attenuation ($A_V\lesssim 1$ mag), 
resolving overestimated luminosities for LRDs in previous studies and alleviating a tension with the so-called So{\l}tan argument. 
This model predicts a distinct spectral evolution as the binary orbit decays through binary-disk interactions and gravitational waves (GWs), 
linking early-stage ``proto-LRD" binaries to the broader AGN population and late-stage ``LRD-descendants" to coalescing binaries detectable in GW experiments.

\end{abstract}

\section{Introduction}

Little red dots (LRDs) represent one of the most intriguing populations uncovered in recent James Webb Space Telescope (JWST) observations.
They are considered to be a new class of broad-line active galactic nuclei (AGN) powered by massive black holes (BHs) with masses of $M_{\rm BH}\simeq 10^{6-8}~\msun$. 
Their spectral energy distribution (SED) is characterized by a distinctive v-shape, with a red continuum in the rest-frame optical and a blue continuum in the rest-frame UV 
\citep{Kocevski_2023,Barro_2024,Matthee_2024,Labbe_2025,Hainline_2025}, and a turnover wavelength near the Balmer limit 
\citep{Furtak_2023a,Greene_2024,Wang_2024b,Setton_2025a}.

The cosmic abundance of LRDs is $\sim 1-2$ orders of magnitude higher than that of UV-bright quasars 
discovered by ground-based telescopes \citep[e.g.,][]{Jiang_2016, Matsuoka_2016, Niida_2020, Matsuoka_2023}, 
yet reaches $\sim 1\% $ of the faint galaxy populations at similar redshifts \citep[e.g.,][]{Kokorev_2024a,Lin_2024,Akins_2025,Kocevski_2025,Lin_2026a}.
If their bolometric luminosities estimated from SED fitting using dust-obscured AGN models are taken at face value,
the inferred cosmic BH accretion rate density (BHAD) appears too high to be reconciled with a standard 10\% radiative efficiency,
suggesting the BHs need to have substantially high spins \citep{Inayoshi_Ichikawa_2024}.
Although rapidly spinning BHs are expected to launch jets via the Blandford-Znajek mechanism \citep{Blandford_Znajek_1977},
LRDs show no obvious (or only weak) X-ray and radio counterparts \citep{Yue_2024, Maiolino_2025X,Mazzolari_2024, Gloudemans_2025}.
This weak non-thermal radiation could be explained by radiative signatures of super-Eddington accreting BHs
\citep{Pacucci_Narayan_2024, Madau_Haardt_2024, Inayoshi_2025a}.
A possible resolution to this tension is that the intrinsic luminosities of LRDs may have been overestimated 
due to uncertainties in dust-reddening corrections in the SED fitting process, as noted by \citet{Inayoshi_Ichikawa_2024}. 
If this is the case, the inferred BHAD could be reconciled with a standard $\sim10\%$ radiative efficiency, 
alleviating the need for extreme BH spin and bringing the LRD population into agreement with the classical 
So{\l}tan–Paczy\'nski argument \citep{Soltan_1982, Yu_Tremaine_2002}.

Previous studies have attempted to explain the v-shape SEDs using two-component models.
These models propose that the UV emission is attributed to an unobscured galaxy or scattered AGN light, and 
the optical emission to an obscured galaxy or obscured AGN \citep{Kocevski_2023,Wang_2024b,Ma_2025a}, 
with a possible contribution from the diversity of dust attenuation laws in UV bands \citep{Z.Li_2025}. 
However, any of these scenarios require fine-tuning the relative contributions of each component to reproduce the observed SED, 
and the underlying physical reason for this ratio remains unclear. 
Whether this ratio is coincidental or reflects a deeper physical process is still debated.
The presence of a Balmer break in observed LRD spectra provides a crucial hint about some contributions of 
stellar light from their host galaxies \citep{Wang_2024b, Kokorev_2024b, Labbe_2024b}.
However, in some cases, the Balmer break is too strong to be explained by stellar populations alone.
Instead, it may result from dense neutral gas absorbers where atomic hydrogen is excited to the $n=2$ state, which 
imprints a Balmer break on the transmitted AGN spectrum \citep{Inayoshi_Maiolino_2025, Ji_2025}.

Furthermore, faint rest-frame near-infrared (NIR) emission of LRDs suggests a lack of hot dust tori surrounding the AGNs
\citep{Williams_2024, Perez-Gonzalez_2024, Wang_2024b,Akins_2025}.
This could be explained by clumpy or extended dusty media \citep{Casey_2024,Z.Li_2025} or simply by lower dust attenuation.
An even more extreme possibility is that the red optical and faint NIR continuum emission can be well fitted by a blackbody spectrum
with a photospheric temperature of $T_{\rm ph} \sim 5000~\K$.
This temperature corresponds to the lowest possible value for a (quasi-)hydrostatic object, set by H$^-$ ions, and follows the so-called Hayashi line
observed as a vertical track on the Hertzsprung–Russell diagram \citep{Hayashi_1961}.
Theoretically, such structures could be maintained as radiation-pressure-dominated envelopes surrounding rapidly accreting BHs at super-Eddington rates.
Potential candidates include hyper-Eddington accretion flows \citep{Inayoshi_Haiman_Ostriker_2016,Takeo_2020,Shi_2023}, quasi-stars 
\citep[e.g.,][]{Begelman_2006,Begelman_2008,Volonteri_2010,Coughlin_2024,Kido_2025}, or
super-massive stars in a tentative stellar phase of direct-collapse BH formation \citep{Hosokawa_2013,Woods_2021,Mayer_2024}.
A recent study by \citet{Naidu_2025} reported an LRD with an exceptionally strong Balmer break,
possibly associated with a highly Compton-thick gas envelope around the AGN, which aligns with the theoretical models.

Beyond their spectral features, the cosmic evolution of LRDs offers further clues to their nature \citep{Kocevski_2025}.
Their abundance rises from $z \gtrsim 8$, peaks at $z \sim 6$, and declines sharply at $z \lesssim 4$.
This transient-like appearance follows a log-normal distribution, suggesting that the AGN phase associated with LRDs is triggered by a distinctive early process and fades over time without recurring at later epochs \citep{Inayoshi_2025b}.
This declining trend is steeper than expected from galaxy (major) mergers alone, which could make LRDs extended objects and 
cause them to lose their distinct characteristics.
Alternatively, \citet{Kahn_2025} proposed that LRDs might transition into elliptical galaxies due to mergers and 
thus lose their characteristics.
If galaxy mergers and subsequent BH coalescences play a role in diminishing LRDs, 
they contribute to gravitational-wave (GW) events at high redshifts \citep[e.g.,][]{Liu_Inayoshi_2025} and the production of a stochastic GW 
background.
A self-consistent theoretical model needs to link this transient behavior to the physical origin of 
their v-shaped SED characterizing LRDs.

In this work, we propose an alternative explanation: the v-shaped SED of LRDs originates from a massive binary BH system, 
where each BH is surrounded by a circum-BH disk (hereafter ``mini-disk") and the system is embedded in a larger circum-binary disk.
In this model, the mini-disks have a higher effective temperature, while the circum-binary disk has a lower effective temperature.
With an appropriate binary separation of $\sim 10^3$ Schwarzschild radii (for the total mass of the BHs), 
the red optical continuum originates from the Wien tail of a blackbody spectrum with $T\sim 5000~\K$ at the inner-edge 
of the circum-binary disk, while the UV continuum is accounted for the mini-disk emission.
Importantly, this model does not require severe dust attenuation, which is often assumed in SED fitting in literature,
but just needs $A_V \lesssim 1$ mag to reproduce the spectral indices used for LRD selection techniques.

This paper is organized as follows.
In Section~\ref{sec:2}, we describe our spectral toy model for a binary BH embedded in a circum-binary disk as well as two mini-disks.
In Section~\ref{sec:3}, we demonstrate how the SED model accounts for the LRD spectral characteristics,
and the spectral evolution led by the binary orbital decay associated with binary-disk interactions and GW emission.
We finally discuss interpretations in our model framework of the LRD abundance and So{\l}tan argument
in Section~\ref{sec:4}, and summarize our conclusions in Section~\ref{sec:5}.

\section{Disk spectral model}\label{sec:2}

\begin{figure}
\centering
\includegraphics[width=75mm]{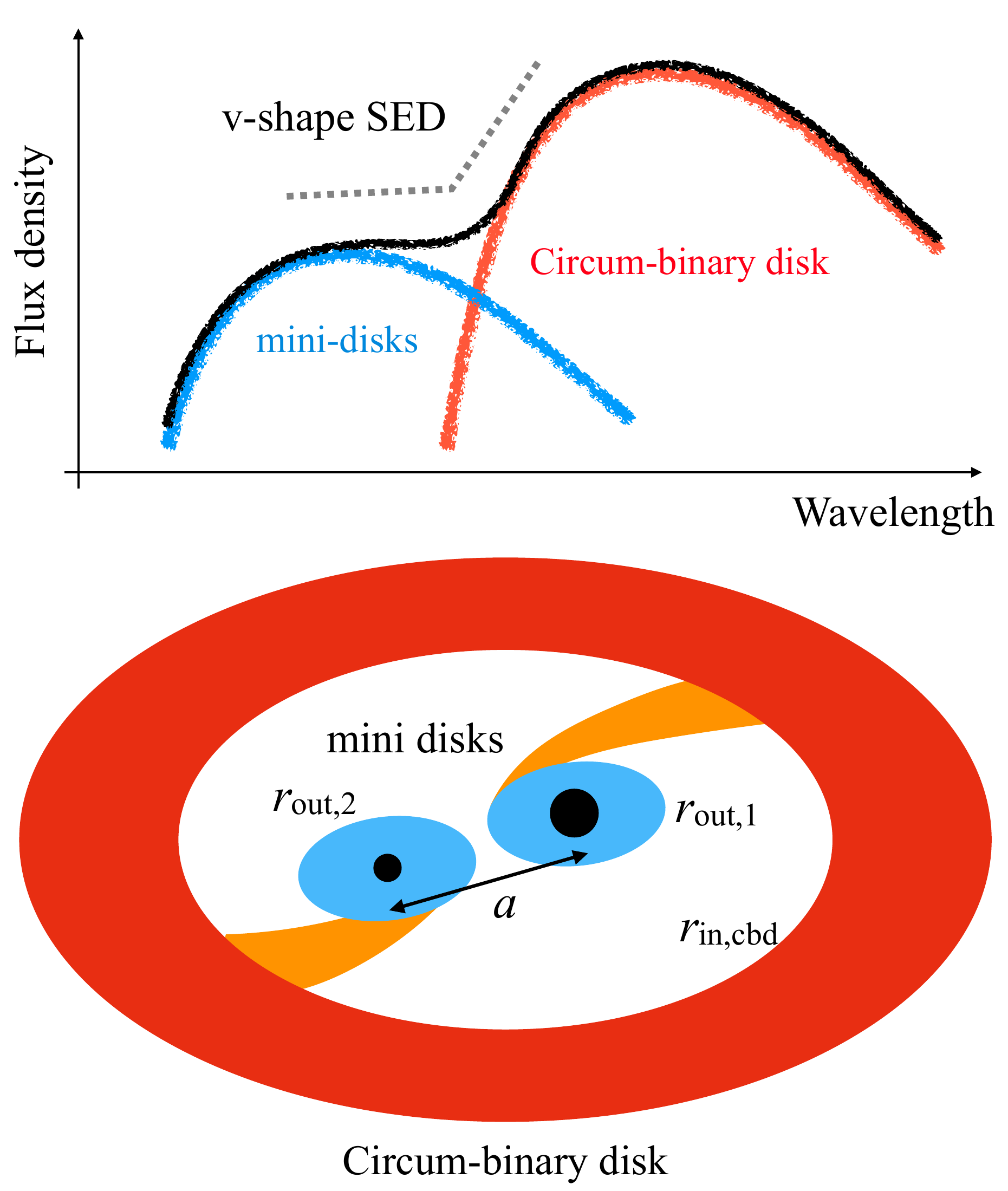}
\caption{A schematic illustration of a binary BH system accreting from the circum-binary disk and mini-disk around each BH.
The expected SED, composed of the colder circum-binary disk and the hotter mini-disks, naturally produces a characteristic v-shape,
consistently observed in LRDs.}
\label{fig:cartoon}
\end{figure}

We describe a spectral model for a binary BH system embedded in a circum-binary disk with each BH surrounded by a mini-disk,
as illustrated in Figure~\ref{fig:cartoon}.
A low-density cavity forms due to binary torques, creating a gap between the inner edge of the circum-binary disk and the outer edges
of the mini disks \citep[e.g.,][]{Artymowicz_Lubow_1994}.
Mass accretion proceeds through the circum-binary disk, along two accretion streams, and onto the two mini-disks \cite[e.g.,][]{MacFadyen_Milosavljevic_2008,
Farris_2014,Farris_2015,Yan_2014,Dorazio+2016,Munoz_2016,Miranda_2017,Tang_2018, Munoz_2019,Westernacher_2022}, 
while mass transfer between the two mini-disks across the Lagrange point is neglected in this work.
Our disk model for the binary system follows previous studies \citep{Roedig_2014,Dorazio+2015}, but we newly incorporate mass-loss 
from the disk system to investigate its effect on the emergent spectrum.

In the standard thin-disk model \citep{Shakura_Sunyaev_1973}, the effective temperature at a distance $r$ from the central BH with a mass of $M$ 
accreting at a rate $\dot{M}$ is given by the energy balance between viscous heating and radiative cooling,
\begin{equation}
T_{\rm eff} = \left[\frac{3}{8\pi \sigma_{\rm SB}} \frac{GM\dot{M}}{r^3}\left(1-\sqrt{\frac{r_{\rm in}}{r}}\right)\right]^{1/4},
\label{eq:Teff_SS}
\end{equation}
where $\sigma_{\rm SB}$ is the Stefan-Boltzmann constant, and $r_{\rm in}$ is the disk inner edge, and the last term inside the parentheses arises from the assumed torque-free inner boundary condition.
In our model, the turnover wavelength of the v-shape SED is set by the temperature contrast between the hottest region in the circum-binary disk and the coldest region in the mini-disks (see Figure~\ref{fig:cartoon}).
To quantify this, we introduce a characteristic temperature $T_0$ defined at $r=a\sim 10^3~R_{\rm S}(\gg r_{\rm in})$ as\footnote{
\citet{Roedig_2014} measure $T_0$ at $a=100~R_{\rm S}$, but their definition of $R_{\rm S}$ should be half of the Schwarzschild radius.}
\begin{equation}
T_0\simeq 6.3\times 10^3 ~\dot{m}^{1/4}M_7^{-1/4}\left(\frac{a}{10^3~R_{\rm S}}\right)^{-3/4}~\K,
\end{equation}
so that the characteristic photon energy falls in the optical bands.
Here, $a$ is the binary semi-major axis, $\dot{m}\equiv \dot{M}/\dot{M}_{\rm Edd}$ is the mass accretion rate normalized by the Eddington value of $\dot{M}_{\rm Edd}\equiv L_{\rm Edd}/\eta_0c^2$, 
$\eta_0=0.1$ is the radiative efficiency, $M_7\equiv M/(10^7~\msun)$, and $R_{\rm S}$ is the Schwarzschild radius of the BH.
The physical quantities related to the BH mass are defined separately for the primary and secondary BH ($M_1\geq M_2 \equiv qM_1$) as well as the total mass ($M=M_1+M_2$).
Recent hydrodynamical simulations suggest that most binaries coupled to their circum-binary disks develop eccentricities of $e\sim 0.5$~\citep{Zrake_2021,DorazioDuffell2021,Siwek_2023}. Here we neglect this complication for simplicity, and assume that binaries are born on orbits with $e\lesssim 0.1$.  In this case, the above simulations find that the orbits are rapidly circularized, and we adopt $e=0$ throughout this study.

%%%%%%%
%   Fig.1   %
%%%%%%%

We note that in the outer region of the accretion disk, where the photospheric temperature is low 
($T_{\rm eff}\ll 5000~\K$), H$^-$ opacity dominates in partially ionized layers, and energy transport in the vertical direction 
is driven by convection \citep{Hayashi_1961,Hoshi_1979,Meyer_Meyer_1982}.
At larger radii, the optical depth eventually drops below unity and the assumption of a locally 
blackbody disk spectrum no longer holds.
In this study, we do not account for these effects in the outermost layers, but instead focus on modeling the SED in 
the UV-optical bands, which can be directly compared to observations.

This temperature $T_0$ corresponds to the photon energy at the turnover point of the v-shape SEDs, giving a characteristic turnover wavelength
\begin{equation}
\lambda_{\rm t}  =\frac{hc}{3kT_0}= 0.76~\dot{m}^{-1/4}M_7^{1/4}\left(\frac{a}{10^3~R_{\rm S}}\right)^{3/4}~\mum .
\label{eq:lambda}
\end{equation}
The numerical factor of 3 in the denominator depends on the width of the gap created by the binary. We adopt this factor of 3 in our analytical estimate of $\lambda_{\rm t}$, while the full SEDs shown in the figures
are obtained by numerically calculating the disk spectrum.
Following \cite{Roedig_2014}, we set the inner edge of the circum-binary disk at $r_{\rm in,cbd}=2a$ and 
the outer edges of the mini-disks at $r_{\rm out,1}=0.27q^{-1/3}a$ for the primary and 
$r_{\rm out,2}=0.27q^{1/3}a$ for the secondary BH. These truncation radii can be estimated analytically \citep{RudakPaczysnki1981,Eggleton_1983} and have been verified in many hydrodynamical simulations \citep[e.g.][]{Artymowicz_Lubow_1994,Farris_2014,Dorazio+2013,Dorazio+2016,Mahesh+2024}.

For a steady-state accretion disk with a constant mass inflow rate ($\dot{M} \propto r^0$), the effective temperature follows 
the radial dependence of $T_{\rm eff}\propto r^{-3/4}$.
This leads to a multi-color blackbody spectrum with a spectral slope of $f_\nu \propto \nu^{\alpha_\nu}$, where $\alpha_\nu = 1/3$.
This blue spectrum ($\alpha_\nu>0$) is a well-known prediction of the standard accretion disk model. 
However, it contradicts the observed spectral index of AGN continua, which is significantly redder, 
with $\alpha_\nu \simeq -0.44$ \citep{VandenBerk_2001}.
The redder spectral index is attributed to multiple physical effects, including electron scattering
being the dominant opacity source over free-free absorption in the inner hot disk, relativistic corrections on both 
the BH gravitational potential and radiative transfer, and possible dust reddening, all of which collectively modify 
the spectrum to resemble observed AGN continua \citep[e.g.,][]{Czerny_1987,Laor_Netzer_1989,Laor_Draine_1993}.

%%%%%%%
%   Fig.2   %
%%%%%%%
\begin{figure}
\centering
\includegraphics[width=84mm]{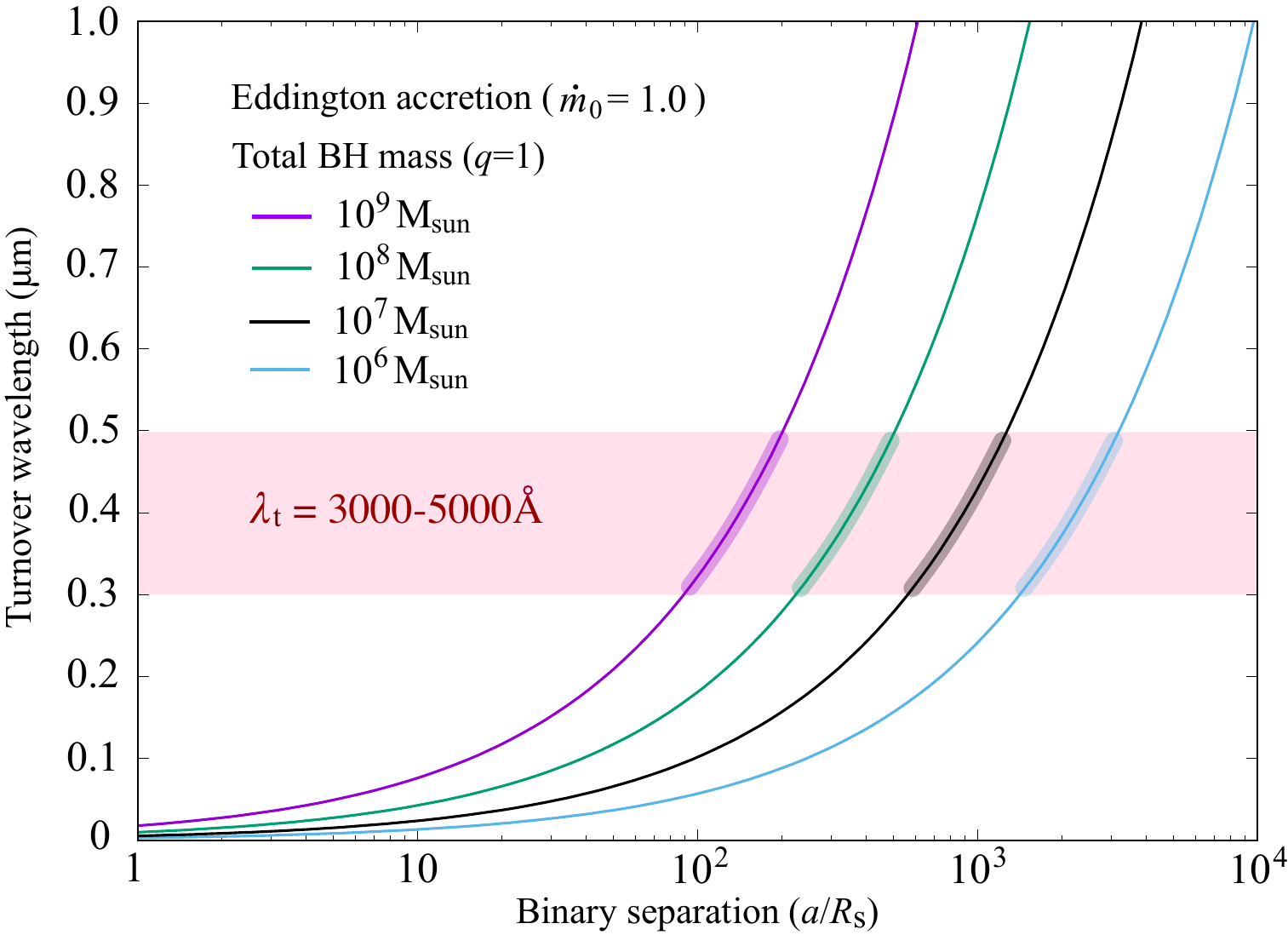}
\caption{Turnover wavelength ($\lambda_{\rm t}$) characterizing the v-shaped LRD spectrum for each model. 
The red shaded region and thick curves present the binary separations where $\lambda_{\rm t}=3000-5000~{\rm \AA}$.}
\label{fig:lambda}
\end{figure}

In this paper, to account for this discrepancy in spectral color, we introduce mass-loss from the accretion flow
parameterized as $\dot{M}\propto r^p$ ($0\leq p<1$).
Under this assumption, the radial dependence of the effective temperature is modified to $T_{\rm eff}\propto r^{-(3-p)/4}$ and 
the spectral slope to $\alpha_\nu =(1-3p)/(3-p)$.
For $p \geq 1/3$, the spectral index becomes redder (i.e., $\alpha_\nu \leq 0$), bringing it closer to observed AGN continuum spectra
\citep{VandenBerk_2001}.
To include this effect, we model the accretion rate as
\begin{equation}
\dot{M}(r) = \dot{M}_0\cdot {\rm max} \left[1,\left(\frac{r}{r_{\rm c}}\right)^p\right],
\end{equation}
where $\dot{M}_0$ is the mass accretion rate at the inner-most disk radius and $\dot{m}_0=\dot{M}_0/\dot{M}_{\rm Edd}$.
We set the transition radius, inside which outflows cease, to $r_{\rm c}=10~R_{\rm S}$.
This choice is somewhat arbitrary, but numerical simulations for BH accretion generally find 
$r_{\rm c}/R_{\rm S}$ to be on the order of $\mathcal{O}(10)$ \citep[e.g.,][]{Abramowicz_2002,Yuan_Narayan_2014}.
Using this prescription in Equation~(\ref{eq:lambda_t}), the turnover wavelength for $p = 0.5$ is
\begin{equation}
\lambda_{\rm t} = 0.43~\dot{m}_0^{-1/4}M_7^{1/4}\left(\frac{a}{10^3~R_{\rm S}}\right)^{5/8}~\mum ,
\label{eq:lambda_t}
\end{equation}
which is close to the Balmer limit wavelength.
In Figure~\ref{fig:lambda}, we show the relationship between the turnover wavelength of the v-shaped LRD spectrum 
for each BH mass when $\dot{m}_0=1$ and $p=0.5$ are adopted. 
The red shaded region and thick curves present the binary separations where $\lambda_{\rm t}=3000-5000~{\rm \AA}$
\citep{Setton_2025a}.

The presence of dense ($n\gtrsim 10^8~\cc$) and moderately fast ($v_{\rm out} \sim 100~\kms$) outflows in LRDs 
has been inferred from spectral analyses of absorption features on top of broad Balmer emission lines \citep{Inayoshi_Maiolino_2025,Rusakov_2025}.
The high density is required to populate atomic hydrogen in the $n=2$ state via collisional excitation within the outflow,
while the outflow velocity is measured from the P Cygni absorption/emission profiles observed in their spectral data.
The inferred outflow rate is substantially super-Eddington, consistent with numerical predictions for outflows launched 
from super-Eddington accreting BHs \citep[e.g.,][]{Hu_2022a}.
Moreover, the weakness of X-ray emission of most LRDs \citep[e.g.,][]{Maiolino_2025X,Akins_2025} and 
the double-sided exponential profiles of Balmer emission lines \citep{Rusakov_2025}
suggest that the central AGNs powering LRDs are deeply embedded in high column density media, consistent
with super-Eddington inflow/outflow systems \citep[e.g.,][]{Madau_Haardt_2024,Inayoshi_2025a,Naidu_2025,deGraaff_2025}.

%%%%%%%
%   Fig.3   %
%%%%%%%
\begin{figure}
\centering
\includegraphics[width=85mm]{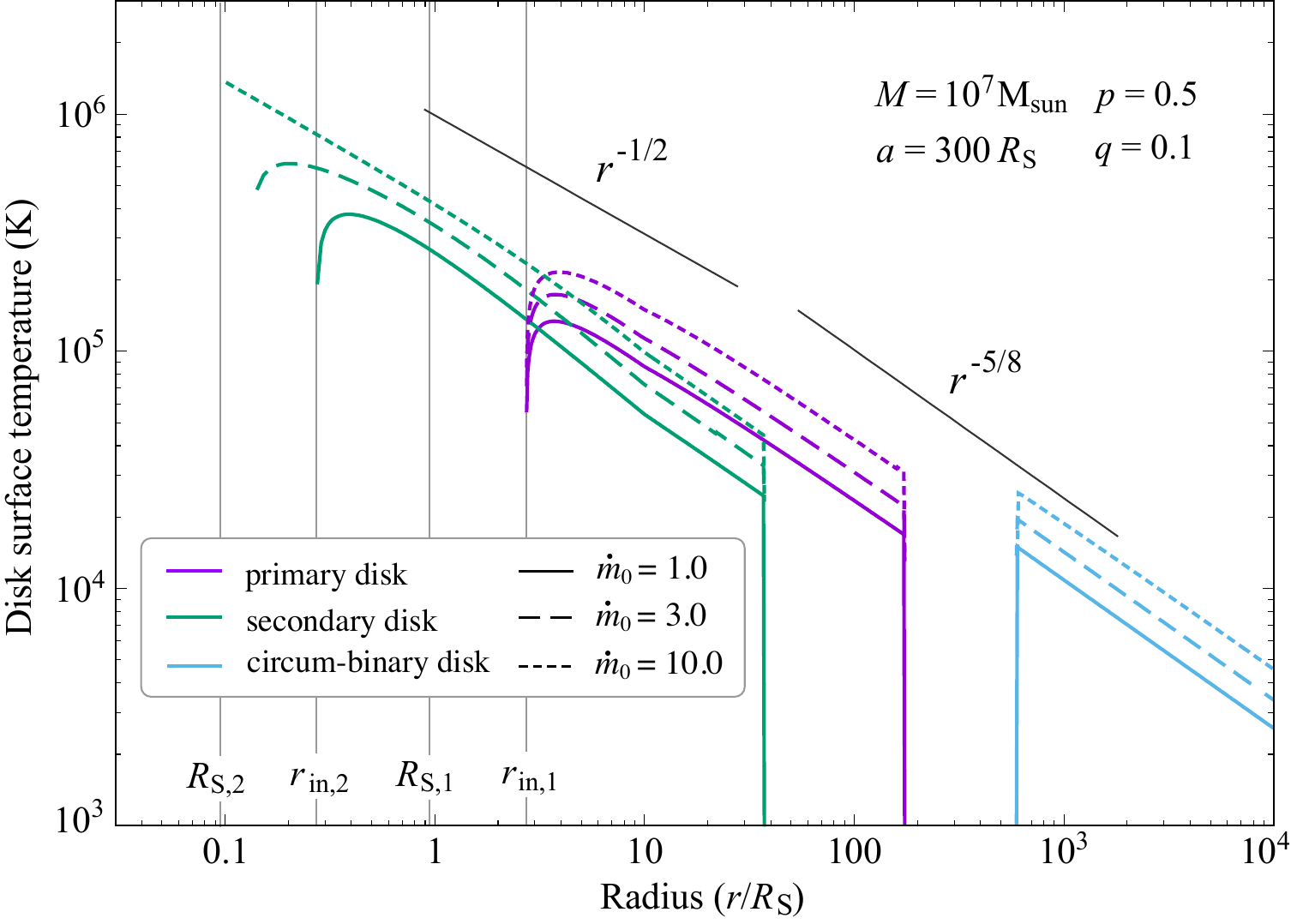}
\caption{Radial structures of the surface temperature of accretion disks around a binary BH with a total mass of $M=10^7~\msun$,
mass ratio $q=0.1$, and orbital separation $a=300~R_{\rm S}$, for three components: the primary mini-disk (purple), 
secondary mini-disk (green), and circum-binary disk (cyan).
Each component is shown for different mass accretion rates of $\dot{m}_0=1$ (solid), $3$ (dashed), and $10$ (dotted).
For the mini-disks, the horizontal axis denotes the radial distance from each BH, normalized by the Schwarzschild radius of the total BH mass. 
For $1\leq \dot{m}_0 \leq 10$, the temperature profiles of both the circum-binary disk and the primary mini-disk follow 
$T_{\rm eff} \propto r^{-5/8}$, consistent with a thin disk model including mass loss characterized by $p = 0.5$.
The secondary mini-disk follows $T_{\rm eff} \propto r^{-1/2}$ in the slim disk regime 
where advection cooling dominates radiative cooling. 
Its inner-edge shifts depending on the accretion rate (see text).
}
\label{fig:Teff}
\end{figure}

The mass accretion rate through the circum-binary disk is divided into two mini-disks \citep[e.g.,][]{Lai_Munoz_2023}.
The ratio of the mass accretion rates generally depends on the mass ratio; namely, $\dot{M}_i = f_i(q) \dot{M}(r)$ for $i=1,2$
with $f_1+f_2=1$.
The tendency observed in numerical simulations is that the secondary BH, which orbits at a larger distance from 
the center of motion accretes more ($f_1<f_2$).
In this work, we adopt $q$-dependent ratios given by
\begin{equation}
f_1=\frac{1}{2}-\frac{4}{9}(1-q),
\label{eq:fmdot}
\end{equation}
and $f_2=1-f_1$ \citep{Munoz_2020,Lai_Munoz_2023}.
This formula fits numerical simulation results for $0.1\lesssim q\leq 1$, and is broadly consistent with
other studies \citep[e.g.,][]{Farris_2014,Duffell+2020,Siwek_2023}.
Using the prescription given in Equation~(\ref{eq:fmdot}) for unequal-mass binaries with $q\simeq 0.1$, 
the primary BH accretes at $\simeq 10\%$ of the total feeding rate, corresponding to an Eddington ratio 
of $\dot{M}_{1}/\dot{M}_{\rm Edd,1}=0.11~\dot{m}_0$.
We note that the assumption of an optically-thick disk is valid only for $\dot{m}_0\gtrsim 0.1$,
above which the accretion flow remains in the standard disk model \citep{Shakura_Sunyaev_1973}.
Below this threshold, the flow is expected to transition to an optically-thin, advection-dominated 
accretion flow \citep[e.g.,][]{Yuan_Narayan_2014}.
In this sub-Eddington regime, the big blue bump seen in AGN UV/optical spectra disappears \citep{Ho_1999}, and 
the accretion energy is instead released as X-ray and radio emission \citep{Ho_2002,Ho_2008}, which are not observed in LRDs
\citep[e.g.,][]{Maiolino_2025X}. This suggests that LRDs cannot be associated with binaries with very unequal ($q\ll 0.1$) masses; however, these binaries are expected to be a small minority produced in galaxy mergers~\citep[e.g.][]{Kelley+2017}.

For mass accretion near or moderately exceeding the Eddington limit ($\dot{m}_0\gtrsim 1$), the temperature 
structure of the mini-disks is characterized by a geometrically-thick, slim disk model.
Within the photon-trapping radius, defined as $r_{\rm tr}/R_{{\rm S},i}\lesssim \dot{M}_i(r_{\rm tr})/\dot{M}_{{\rm Edd},i}$, 
the heat generated by viscosity is advected inward with dense inflows rather than being transported by radiative diffusion
\citep{Abramowicz_1988,Kato_2008}.
The radial profile of the disk surface temperature is given by \cite{Watarai_2006} as
\begin{equation}
T_{\rm eff}=7.91\times 10^6f^{1/8} M_{i,7}^{-1/4}\mathcal{F}(r,\dot{m}_i)\left(\frac{r}{R_{{\rm S},i}}\right)^{-1/2}~\K,
\label{eq:slim}
\end{equation}
where $f$ is the ratio of the advection cooling rate to the viscous heating rate\footnote{The fitting form of the ratio is given by \cite{Watarai_2006} as $f=0.5(X^2+2-X \sqrt{X^2+2})$, where $X=0.281r/(\dot{m} R_{\rm S})$.}, 
and the function $\mathcal{F}$ characterizes the disk-inner boundary as 
$\mathcal{F}(r,\dot{m}_i) = [1-\sqrt{r_{\rm in}/r}]^{1/4}$ for $\dot{m}_i<4$, 
and $\mathcal{F}(r,\dot{m}_i)=1$ for $\dot{m}_i \geq4$.
The radius of the disk inner edge ($r_{\rm in}$) is set to the inner-most stable circular orbit (ISCO) radius 
for a non-spinning BH ($r_{\rm in}=3~R_{{\rm S},i}$) 
for $\dot{m}_i \leq 1$, and for $1<\dot{m}_i <4$ is calculated with a linear interpolation in the plane of 
$\log \dot{m}_i - \log r_{\rm in}$ between the ISCO radius and $1.1~R_{{\rm S},i}$.
We note that the surface temperature profile reproduces the conventional form of $T_{\rm eff} \propto r^{-(3-p)/4}$ 
for the sub-Eddington regime (see Equation~\ref{eq:Teff_SS}).

%%%%%%%
%   Fig.4   %
%%%%%%%
\begin{figure*}
\centering
\includegraphics[width=85mm]{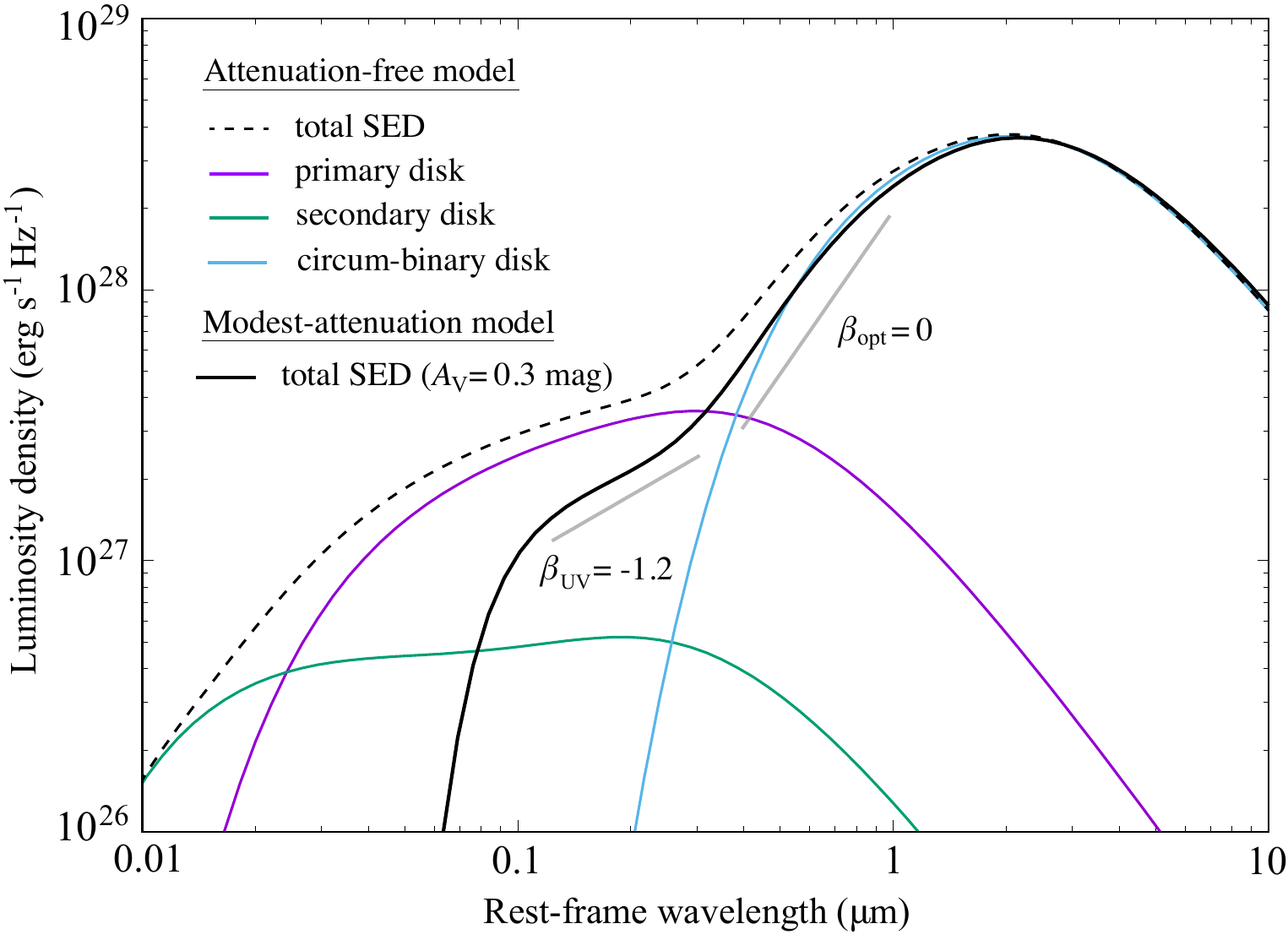}\hspace{5mm}
\includegraphics[width=85mm]{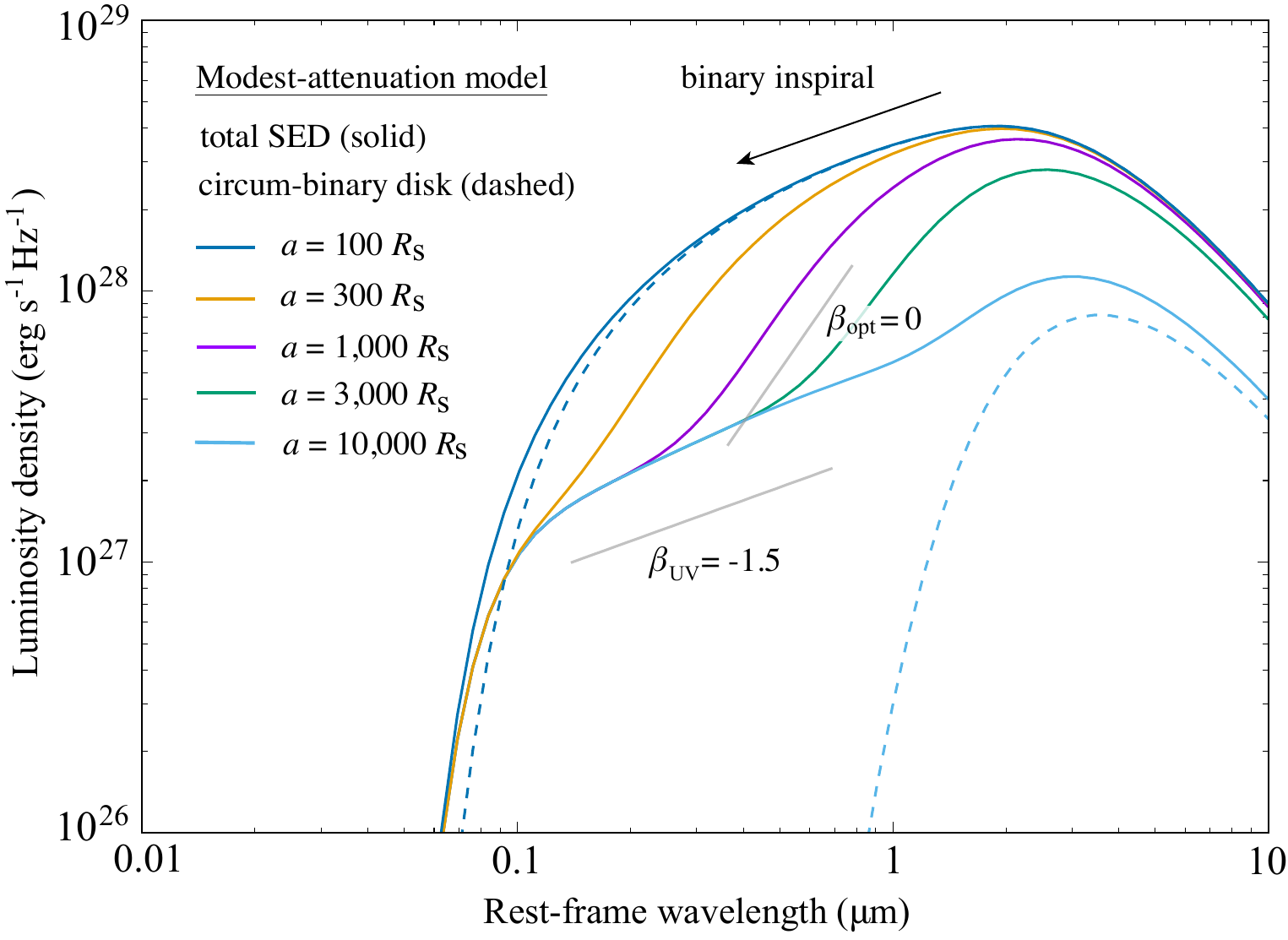}
\caption{Spectra of a binary BH system with $M=10^7~\msun$, $q=0.1$, $\dot{m}_0=1.0$, and $p=0.5$. {\it Left}: The case with $a=10^3~R_{\rm S}$. 
The attenuation-free SED for the primary mini-disk (purple), secondary mini-disk (green), circum-binary disk (cyan), and the total system (black dashed). 
The black solid curve shows the SED with modest attenuation ($A_V=0.3$ mag), yielding UV and optical slopes of $\beta_{\rm UV}= -1.2$ and $\beta_{\rm opt}= 0$ (gray lines),
consistent with the color selection criteria for LRDs.
{\it Right}: SEDs for various binary separations in the range over $a=10^2-10^4~R_{\rm S}$ (solid curves). 
For reference, the dashed curves present the circum-binary disk spectra for the smallest and largest separations.
As the orbital separation shrinks, the surface temperature of the circum-binary disk increases, shifting the peak wavelength toward shorter values.}
\label{fig:spectrum}
\vspace{3mm}
\end{figure*}

Figure~\ref{fig:Teff} illustrates the surface temperature profiles around a binary BH with a total mass of $M=10^7~\msun$,
mass ratio $q=0.1$, orbital separation $a=300~R_{\rm S}$, and mass-loss parameter $p=0.5$,
for three components: the primary mini-disk (purple), secondary mini-disk (green), and circum-binary disk (cyan).
Each disk component is shown for different mass accretion rates of $\dot{m}_0=1$ (solid), $3$ (dashed), and $10$ (dotted).
The slim disk model is also applied to the circum-binary disk, but the temperature structure is well described by the thin disk
model as its inner-edge is located outside the photon-trapping radius.
The secondary mini-disk and the inner-region of the primary mini-disk follow $T_{\rm eff} \propto r^{-1/2}$, 
where advection cooling dominates radiative cooling.

%\vspace{5mm}
\section{Results}\label{sec:3}

\subsection{Parameter dependence}\label{sec:3.1}

Figure~\ref{fig:spectrum} presents the spectra of a binary BH system with $M=10^7~\msun$, $q=0.1$, $\dot{m}_0=1.0$, and $p=0.5$. 
In the left panel, we show the case for a binary separation of $a=10^3~R_{\rm S}$. 
The attenuation-free SED is decomposed into contributions from the primary mini-disk (purple), secondary mini-disk (green), and circum-binary disk (cyan).

The total SED (black dashed) is dominated by the primary mini-disk in the UV regime and the circum-binary disk in the optical regime
with a turnover wavelength of $\lambda_{\rm t}\simeq 0.4~\mum$, consistent with both observations and the analytical estimate 
in Equation~(\ref{eq:lambda_t}).
At this binary separation, the optical continuum arises from the Wien tail of the circum-binary disk inner edge, 
while the UV continuum follows the Rayleigh-Jeans tail of the outer edge of the primary mini-disk.
Although the secondary mini-disk contributes less to the UV luminosity, its spectrum extends to shorter wavelengths at 
$\lesssim 0.02~\mum$ with a bluer slope than the primary disk, due to super-Eddington accretion
modeled with a slim disk (see Equation~\ref{eq:slim}).
This model naturally produces a v-shaped SED even in the absence of dust attenuation, host galaxy stellar light,
and scattered AGN light.

The black solid curve shows the SED with modest attenuation ($A_V=0.3$ mag), which suppresses the UV component.
This model yields UV and optical slopes of $\beta_{\rm UV}= -1.2$ and $\beta_{\rm opt}= 0$ (gray lines),
consistent with the color selection criteria for LRDs \citep{Matthee_2024,Greene_2024,Kocevski_2025}.
A key feature of our model is that only a low level of dust attenuation is needed,
in contrast to most previous studies on LRD SEDs, which postulate significant obscuration with $A_V \sim 3-4$ mag.
This weaker attenuation also aligns with the weak NIR fluxes observed in LRDs with JWST MIRI photometry
\citep[e.g.,][]{Wang_2024a,Wang_2024b,Setton_2025b,Akins_2025}, as the reprocessed IR radiation should be weaker 
if the intrinsic SED is UV-faint and redder in the optical bands.
In Section~\ref{sec:soltan}, we discuss an application of our low-attenuation model to the So{\l}tan argument.

It is worth noting that this spectral model favors a mass ratio of $q\lesssim 0.1$, 
where the primary-disk accretion fraction $f_1$ converges $\simeq 0.055-0.1$ (see Equation~\ref{eq:fmdot}).
Based on the disk energetics argument in Appendix~\ref{sec:AppDisk},
the luminosity density ratio between the UV (from the primary mini-disk) and optical (from the circum-binary disk) 
bands is evaluated in Equation~(\ref{eq:Lnuratio}) as 
\begin{equation}
    \frac{L_{\nu_{\rm uv}}}{L_{\nu_{\rm opt}}}\sim 0.12~\left(\frac{f_1}{0.1}\right)
    \left(\frac{\nu_{\rm opt}}{0.1\nu_{\rm uv}}\right)
    \left(\frac{a}{10^3~R_{\rm S}}\right)^{1/2},
    \label{eq:Luvopt}
\end{equation}
where $\nu_{\rm uv}$ and $\nu_{\rm opt}$ are the characteristic photon frequencies at 
the UV and optical bands, respectively.
Under these conditions, the resulting SED matches the LRD color selection criteria even without or 
with moderate dust reddening (see Figure~\ref{fig:spectrum}).
In contrast, as the mass ratio approaches unity, the luminosity density ratio increases to 
$\sim \mathcal{O}(1)$, leading to a spectrum dominated by the mini-disk.
This case fails to reproduce the characteristic v-shaped SED of LRDs, even when dust reddening is considered.

%%%%%%%
%   Fig.5   %
%%%%%%%
\begin{figure*}
\centering
\includegraphics[width=150mm]{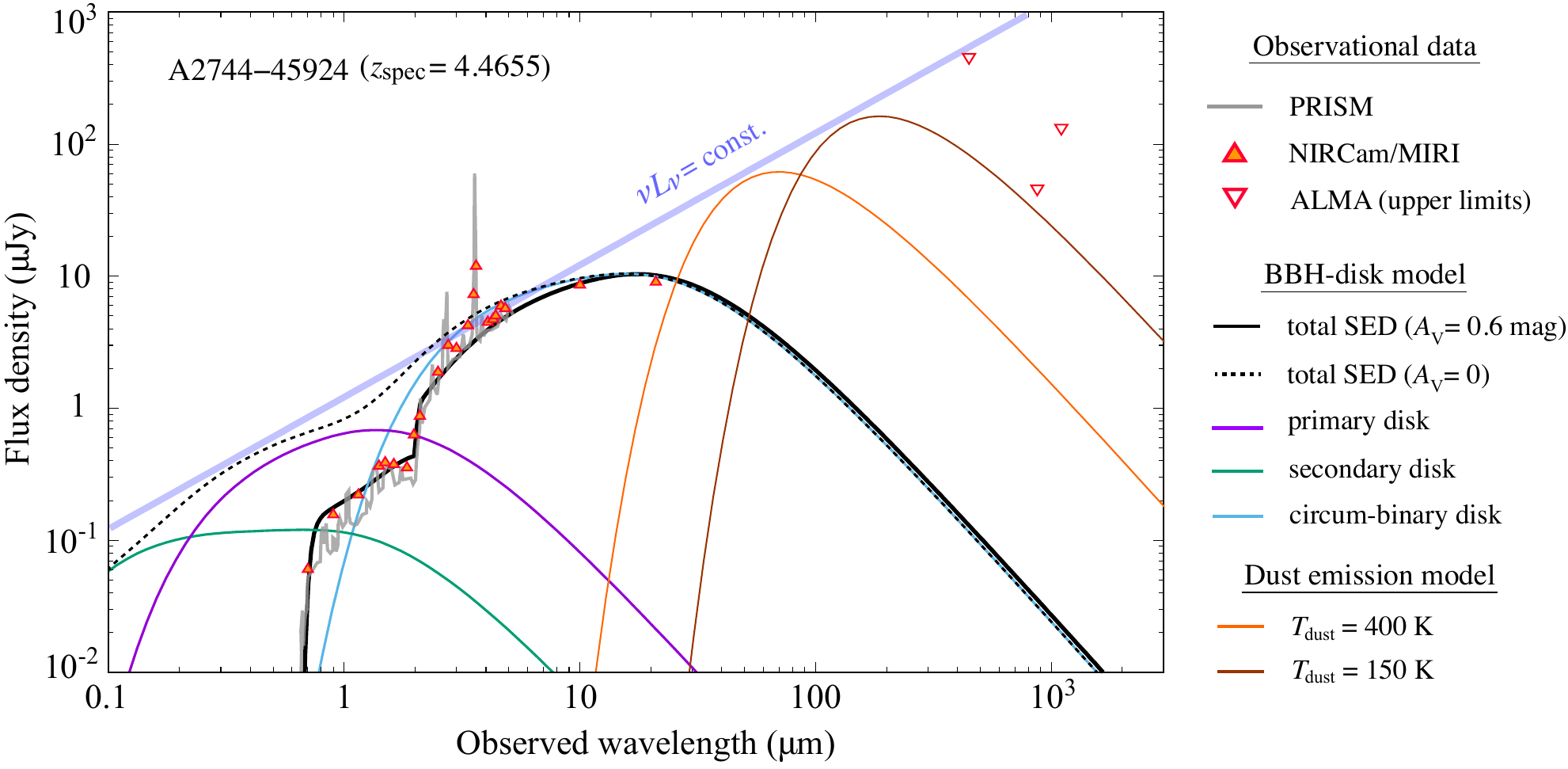}
\caption{A spectral model for the brightest LRD, A2744-45924, at $z_{\rm spec}=4.46$ with NIRCam photometry and PRISM data \citep{Labbe_2024b}
and MIRI photometry and ALMA flux upper limits \citep{Setton_2025b}.
The binary BH-disk SED model assumes parameters $M=2\times 10^8~\msun$, $q=0.1$, $\dot{m}_0=3.0$, $p=0.5$, $a=300~R_{\rm S}$, and $A_V=0.6$ mag,
and reproduces the observed SED with only moderate attenuation, as the Wien-tail spectrum of the circum-binary disk is intrinsically red in the rest-frame optical bands.
Blackbody spectra with dust temperatures of $T_{\rm dust} = 400~\K$ (orange) and $150~\K$ (brown) are overlaid,
with normalizations keeping photon energy conservation.
Given that the total IR luminosity is limited to $\sim 2\times 10^{11}~\lsun$, the IR spectrum of warm dust emission is compatible with the MIRI and ALMA observations.
}
\label{fig:A2744}
\vspace{3mm}
\end{figure*}

In the right panel of Figure~\ref{fig:spectrum}, we show how the SED evolves as the binary separation decreases 
from $a=10^4~R_{\rm S}$ to $100~R_{\rm S}$ (solid curves).
For reference, the dashed curves present the circum-binary disk spectra for the largest and smallest separations.
For $a=10^4~R_{\rm S}$, the SED at $\lambda \lesssim 1~\mum$ is dominated by the primary mini-disk 
with a red spectral slope of $\alpha_\nu =-1/5$ ($p=0.5$).
As the binary inspirals, the circum-binary disk heats up, increasing its luminosity and shifting the peak flux density 
to shorter wavelengths.
By the time the binary reaches $a=100~R_{\rm S}$, the circum-binary disk fully dominates the SED and 
the characteristic v-shaped SED fades, leaving only a red continuum.
This spectral evolution in the emergence of a v-shaped SED is observed for different BH masses 
though the binary separation differs (see Figure~\ref{fig:lambda}).

However, there is a caveat for interpreting this figure.
Once the binary separation shrinks below $a\lesssim 600~R_{\rm S}$, the orbital decay begins to accelerate 
due to the energy and angular momentum loss via GW radiation.
At later stages, the inner-edge of the circum-binary disk can no longer follow the rapidly inspiraling binary 
but instead approximately freezes at the decoupling radius on a viscous timescale (\citealt{Dittmann+2023,Krauth_2023,O'Neill_2025}; see more discussion in Section~\ref{sec:evol}).
Thus, the case labeled $a = 100~R_{\rm S}$ should be interpreted as the stage when the inner edge of 
the circum-binary disk is located at $r_{\rm in,cbd} = 200~R_{\rm S}$, rather than the moment when the binary 
reaches $a = 100~R_{\rm S}$.

Numerical simulation studies have shown that binary BH systems produce SEDs with a characteristic dip at the wavelength where
the spectrum of the hot mini-disks intersects with that of the colder circum-binary disk \citep[e.g.,][]{Roedig_2014, Farris_2015, Tang_2018}.
In these simulations, the binary separation is typically assumed to be $a\ll 10^3~R_{\rm S}$, imprinting its dip feature at high photon energies 
in the UV to X-ray bands.
The spectral turnover energy (or wavelength) is described by the Equation~(\ref{eq:lambda}) in the absence of mass loss ($p=0$).
These studies also demonstrate that shock heating powered by the binary orbital energy can significantly enhance the total luminosity and 
harden the spectrum, often producing strong UV and X-ray features (see also \citealt{Westernacher_2022} and \citealt{Krauth_2023}, as well as an
earlier study by \citealt{Lodato+2009} proposing that tidal heating of the cavity wall can produce similar effects).
However, such high-energy components are generally absent in the redder SEDs of LRDs \citep[e.g.,][]{Lambrides_2024}. Notably, they are also absent in follow-up X-ray observations of optically identified massive binary candidates with both Chandra \citep{Saade+2020} and XMM-Newton \citep{Saade+2024}.
The hard UV/X-ray features can be suppressed in binaries if mini-disks are weak or absent. 
Recent work suggests that this is indeed the case when the accretion flow is cold \citep{Tiede_2025},
offering a possible pathway to reconcile binary BH scenarios with LRD observations.
Thus far, such high-energy emission from shock-heated accretion streams and mini-disks remains theoretically uncertain
because it depends sensitively on the detailed disk thermodynamics and morphology as well as the binary orbital dynamics. 
Given these uncertainties, we do not explicitly include these components in our model, but instead focus on the more robust emission 
from the circum-binary disk, which dominates the red optical continua observed in LRD spectra and is less affected by the complex inner-disk physics.

\subsection{Case studies}

Figure~\ref{fig:A2744} demonstrates that our binary BH-disk model explains the SED of the brightest LRD \citep[A2744-45924,][]{Labbe_2024b}.
A2744-45924 has been confirmed as a broad-line AGN at $z_{\rm spec}=4.46$ with a Balmer break feature detected in its PRISM spectrum \citep{Greene_2024}.
Follow-up observations have revealed the properties of its emission lines \citep{Labbe_2025} and dust emission through MIRI and ALMA observations \citep{Setton_2025b}.
Spectral fitting by \citet{Labbe_2024b} indicates strong attenuation ($A_V = 2.4 - 3.0$ mag), leading to an estimated BH mass of $M_{\rm BH} = 7 \times 10^8~\msun$
via the single-epoch method based on broad H$\alpha$ emission.

We present a binary BH-disk SED model with parameters $M=2\times 10^8~\msun$, $q=0.1$, $\dot{m}_0=3.0$, $p=0.5$, and $a=300~R_{\rm S}$.
This model reproduces the observed SED of A2744-45924 with only moderate attenuation ($A_V=0.6$ mag), 
as the Wien-tail spectrum of the circum-binary disk is intrinsically red in the rest-frame optical bands.
The Balmer-break feature in the model SED is added by using the analytical form based on the AGN transmitted spectrum
with hydrogen column density of $N_{\rm H}=10^{23}~{\rm cm}^{-2}$
\citep[][see more details in Appendix~\ref{sec:AppBalmer}]{Inayoshi_Maiolino_2025}.
The orbital separation of $a=300~R_{\rm S}$ aligns with the values highlighted in Figure~\ref{fig:lambda}, ensuring the v-shaped SED with a turnover near the Balmer limit.
Although \citet{Labbe_2024b} estimate the BH mass as $M_{\rm BH}=7\times 10^8~\msun$ assuming $A_V=2.4-3.0$, 
our model suggests a lower total BH mass $M_{\rm BH}=2\times 10^8~\msun$ but with a modest super-Eddington accretion 
rate of $\dot{m}_0=3$.
This combination maintains the observed optical luminosity while ensuring that the turnover wavelength $\lambda_{\rm t}$
remains within the LRD selection window. 
Importantly, the observed flux density is not treated as a free parameter but rather emerge self-consistently from 
the intrinsic energetics of the accretion disks.
We note that these parameters reproducing the observed spectrum of A2744-45924 are not obtained from a fitting procedure.
Instead, this work focuses on the model prescription and its physical implications, while the details of a dedicated fitting methodology 
are left to future work.

Our binary BH-disk model can also explain the SED of another recently reported LRD, MoM-BH$^\ast$-1 \citep{Naidu_2025}.
This object shows the strongest Balmer break among all spectroscopically observed LRDs.
This feature cannot be explained by stellar populations alone but likely arises from absorption by 
dense gas surrounding the nuclear BH \citep[e.g.,][]{Inayoshi_Maiolino_2025,Ji_2025}.
\citet{Naidu_2025} suggest that MoM-BH$^\ast$-1 represents a super-Eddington accreting BH embedded 
in an extremely dense, possibly dust-poor gas environment.
Such a surrounding medium could resemble a quasi-star \citep{Begelman_2006, Begelman_2008} or 
hyper-Eddington accretion flows \citep{Inayoshi_Haiman_Ostriker_2016,Takeo_2020}, characterized by 
a sufficiently low photospheric temperature of $T_{\rm ph}\sim 5000-7000~\K$ \citep{Hayashi_1961}.
These interpretations align with our model, where the inner edge of the circum-binary disk produces 
a similar surface temperature (see also a BH-envelope model proposed in \citealt{Kido_2025}).

A key advantage of our binary BH-disk model is that only modest attenuation is required to reproduce the observed SED.
This moderate reddening is further supported by the recent discovery of several LRDs showing a prominent Balmer break (e.g., \citealt{Naidu_2025,deGraaff_2025,Taylor_2025_z9.3}).
This, in turn, implies that the intrinsic rest-frame UV emission is not particularly strong,
resulting in only moderate NIR flux from reprocessed dust emission.
To illustrate this, we overlay blackbody spectra with dust temperatures of $T_{\rm dust} = 400~\K$ (orange) and $150~\K$ (brown),
with normalizations keeping photon energy conservation.
Given that the total IR luminosity is limited to $\sim 2\times 10^{11}~\lsun$, the allowed dust temperature range is $T_{\rm dust} \sim 150-400~\K$.
In contrast, if the intrinsic UV emission were significantly brighter, the range of dust temperatures required to match MIRI and ALMA constraints would be much narrower.
This result is crucial, as a significant fraction of LRDs exhibit faint rest-frame NIR (MIRI bands) and FIR (ALMA bands) fluxes,
suggesting the absence of hot dusty tori in their nuclei and minimal ongoing star formation.

\subsection{Evolution of binary BHs and their SEDs}\label{sec:evol}

\begin{figure}
\centering
\includegraphics[width=83mm]{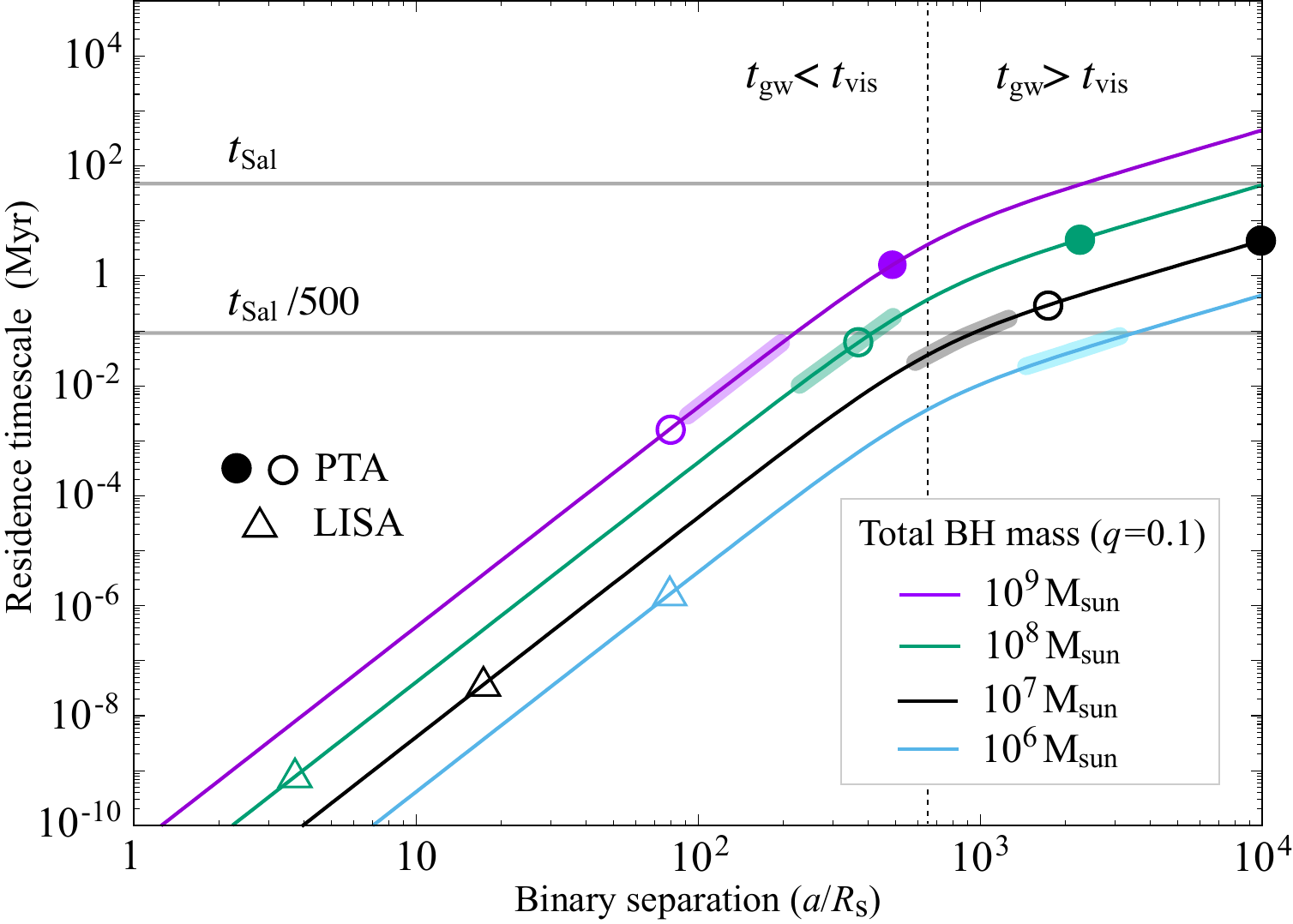}
\caption{Binary residence time as a function of binary separation for different total BH masses: $M_{\rm BH}=10^6~\msun$ (cyan), $10^7~\msun$ (black),
$10^8~\msun$ (green), and $10^9~\msun$ (purple). Equal-mass ratio BHs accreting at the Eddington rate for the total mass are assumed.
For each curve, the filled circle, open circle, and triangle mark the epochs when the GW frequency reaches
$f_{\rm gw}=2~{\rm nHz}$, $32~{\rm nHz}~(=1~{\rm yr}^{-1})$, and $0.032~{\rm mHz}$, corresponding to the frequency ranges probed by the PTA and LISA. 
The vertical line indicates the transition where binary orbital decay becomes dominated by GW emission rather than interactions with the gaseous disk.
The thick curves present the binary separations where $\lambda_{\rm t}=3000-5000~{\rm \AA}$. 
}
\label{fig:time}
\vspace{0mm}
\end{figure}

In our scenario, the binary orbital evolution governs the spectral evolution.
Figure~\ref{fig:time} summarizes the key timescales and characteristic binary separations for various total BH masses.

At wider separations, the binary orbital evolution is primarily governed by interactions with the surrounding gas disk.
If the secondary BH opens a gap in the disk, its evolution proceeds via type-II migration on a viscous timescale 
\citep[e.g.,][]{Lin_Papaloizou_1979,Lin_Papaloizou_1986,Armitage_Natarajan_2002}.
\begin{align}
t_{\rm vis} & = \frac{1}{\alpha \Omega}\left(\frac{a}{H}\right)^2
\simeq 0.14~{\rm Myr}~\alpha_{0.01}^{-1}M_7\left(\frac{a}{10^3~R_{\rm S}}\right)^{3/2},
\end{align}
where we assume a geometrically-thin disk with a scale-height of $H/a=0.01$ and a $\alpha$-viscous parameter
of $\alpha=0.01$, consistent with magneto-hydrodynamic simulations 
\citep[e.g.,][]{Balbus_Hawley_1998,Stone_Pringle_2001,Bai_2011}.
\citet{Haiman_2009} provide a more detailed analytical treatment of the disk structures and orbital decay within AGN,
but our simple approach with a constant $H/a$ ratio captures the overall trend.

As the binary separation shrinks, GW emission takes over driving the orbital decay through its orbital energy 
and angular momentum loss on a timescale:
\begin{equation}
t_{\rm gw} \simeq 0.14~{\rm Myr}~q_s^{-1} M_7 \left(\frac{a}{10^3~R_{\rm S}}\right)^4,
\label{eq:gw}
\end{equation}
\citep[e.g.,][]{Peter_Mathews_1963}, where $q_s\equiv 4q/(1+q)^2$ is the symmetric mass ratio.
Figure~\ref{fig:time} presents the residence timescale defined as $t_{\rm res}^{-1}\equiv t_{\rm vis}^{-1}+t_{\rm gw}^{-1}$
(solid curves).

By equating the GW and viscous timescales ($t_{\rm gw}=t_{\rm vis}$), we obtain the critical orbital separation 
at which GW emission begins to dominate the binary inspiral evolution
\begin{equation}
a_{\rm gw}^{\rm vis} \simeq 6.5\times 10^2~q_{s0}^{2/5} \alpha_{0.01}^{-2/5}~R_{\rm S},
\end{equation}
where $q_{s0}=q_s/0.33$ is the symmetric mass ratio normalized by the value for $q=0.1$.
The corresponding GW frequency (i.e., twice the orbital frequency) in the source frame is calculated by
\begin{equation}
f_{\rm gw} \simeq 71.8~{\rm nHz}~M_7^{-1}\left(\frac{a}{10^3~R_{\rm S}}\right)^{-3/2}.
\label{eq:fgw}
\end{equation}
For lower-mass BHs ($M_{\rm BH}\lesssim 10^7~\msun$), the binary enters the frequency range detectable
by the pulsar timing array (PTA) experiments (filled and open circles in Figure~\ref{fig:time}),
while its orbital evolution is still driven by disk viscosity.
As the binary inspirals further and exits the PTA band, the characteristic v-shaped SED feature emerges
(the shaded region on each curve in Figure~\ref{fig:time}).
In contrast, for more massive BHs ($M_{\rm BH}\gtrsim 5\times 10^7~\msun$), the binary remains within the PTA-detectable 
GW frequency range even when the LRD signature becomes observable.
We note that the separation range where the GW and LRD signatures coexist extends further 
for high-redshift sources, as the circle symbols shift by a factor of $\sim 3~[(1+z)/6]^{-2/3}$.
Eventually, these inspiraling binaries enter the frequency band at $f_{\rm gw}\simeq 0.032~{\rm mHz}$ (triangle symbols), 
corresponding to the ranges probed by the Laser Interferometer Space Antenna (LISA).

The duration of the LRD phase, when the spectral properties match the color selection criteria, corresponds to only
$\sim 0.2\%$ of the Eddington-limited accretion $e$-folding timescale (i.e., the Salpeter timescale, $t_{\rm Sal}$).
Taking the Salpeter timescale as a reference for the AGN lifetime, the duty cycle of LRD spectral features increases 
when the viscous timescale is longer.
This can occur due to inefficient turbulent viscosity ($\alpha<0.01$) and/or even thinner disk 
configurations ($H/a<0.01$). 
This suggests that the majority of accretion activity for these BHs occurs during normal AGN phases, when the binary separation 
is larger and the residence time is longer. 
As shown in Figure~\ref{fig:spectrum}, the SED shape at separations of $a\sim 10^4~R_{\rm S}$ resembles that of the typical 
unobscured AGNs. A more quantitative discussion of the abundance of this `proto-LRD' population is presented in Section~\ref{sec:4.1}.

%%%%%%%
%   Fig.6   %
%%%%%%%

It is worth noting that the dynamics of the circum-binary disk decouples from the binary evolution when $a \ll a_{\rm gw}^{\rm vis}$.
At this stage, GW radiation rapidly drives the binary inspiral, while the inner edge of the circum-binary disk remains 
nearly stationary, unable to adjust on such a short timescale.
As a result, the characteristic red optical spectrum originating from the circum-binary disk persists for 
a duration of $t_{\rm vis}\simeq 74~M_7~{\rm kyr}$.
Meanwhile, the spectra of the two mini-disks evolve much faster on the GW-driven inspiral timescale.
For instance, at $a\lesssim 0.4~a_{\rm gw}^{\rm vis}\sim 260~R_{\rm S}$, the GW timescale is an order of magnitude shorter 
than the viscous timescale.

In this decoupling regime, strong binary torques have been proposed to compress the disk inside the secondary orbit \citep{Armitage_Natarajan_2002}.
This GW-driven disk squeezing triggers rapid accretion onto the primary BH and potentially produces a luminous precursor event
\citep[e.g.,][]{Chang_2010,Cerioli_2016,Fontecilla_2017} $-$ days to hours before the BH merger in the source frame.
If the squeezing process continues until the binary reaches $a\simeq 20~R_{\rm S,1}$, the inner-mini disk becomes
geometrically thick ($H/a\sim 1$) and envelops the compact binary system within a nearly spherical structure \citep{Fontecilla_2017}.
By this stage, the accretion rate onto the primary BH increases significantly, scaling as $t_{\rm vis}/t_{\rm gw}\sim \mathcal{O}(100)$.
However, due to efficient photon trapping, the increase in radiative luminosity is limited to a factor of a few \citep[e.g.,][]{Abramowicz_1988,Ohsuga_2005,Jiang_2014,Sadowski_2015}.
On the other hand, recent high-resolution simulations suggest an alternative picture in which the mini-disks are tidally disrupted 
and depleted prior to merger, rather than squeezed \citep{Krauth_2023}. 
In this scenario, the UV/X-ray emission drops by four orders of magnitude as the binary evolves\footnote{A recent GRMHD simulation~\citep{Ennoggi_2025} finds a similar destruction of the minidisks just prior to merger, but does not find a drop in the total bolometric luminosity, which is dominated by a large Poynting flux.}.
These contrasting outcomes highlight the theoretical uncertainty regarding the final stages of binary-disk interaction.

Finally, following the merger, the remnant BH receives a recoil kick opposite to the direction of GW emission.
This motion induces prompt shocks between the disk material bound to the recoiling remnant and the outer disk,
potentially generating an observable electromagnetic counterpart \citep{Lippai_2008,SchnittmanKrolik2008, Megevand+2009,Megevand+2010,Corrales+2010,Rossi+2010}.

\section{Discussion}\label{sec:4}

\subsection{Abundance}\label{sec:4.1}
The high abundance of LRDs in a given cosmic volume is an intriguing aspect of this population, 
with a comoving number density of $\phi \simeq 10^{-5}-10^{-4}~{\rm Mpc}^{-3}$.
This is approximately $\sim 1-2$ orders of magnitude higher than that of UV bright quasars, 
but occupies $\lesssim 10\%$ of unobscured broad-line AGNs \citep{Harikane_2023_agn,Maiolino_2024_JADES,Taylor_2025}
and possibly $\sim 1\%$ of obscured Type 2 AGNs (which lack broad emission lines, and are selected based on narrow line emission diagnostics; \citealt{Scholtz_2025}) in the same redshift range.

In our scenario, the spectral features of LRDs originate from binary BHs at orbital separations of $a\simeq 10^3~R_{\rm S}$.
However, the majority of accretion activity for these BHs occurs in normal AGN phases, when the binary separation is wider and thus the residence time is longer.
At separations of $10^3\lesssim a/R_{\rm S}\lesssim 10^4$, the SED\footnote{The rest-frame UV-to-optical emission is primarily produced by the two mini-disks, as shown in Figure~\ref{fig:spectrum}. At these large separations, the surface temperature of the circum-binary disk is not determined by radiative cooling but by convective heat transport \citep{Hoshi_1979,Mineshige_Osaki_1983}. As the binary separation increases further, the circum-binary disk becomes optically thin. We do not model this regime in this work.} resembles that of typical AGNs and does not show the distinctive v-shape seen in LRDs, while retaining a similar UV magnitude (see the right panel of Figure~\ref{fig:spectrum}).
In this regime, binaries are coupled well to the circum-binary disk, and thus their residence time is expected to be governed roughly 
by the viscous time $t_{\rm vis}(\propto a^{3/2})$.
Given the observed abundance ratio between LRDs and unobscured AGNs with comparable UV magnitudes, a factor of $\sim 30$ difference in number density \citep[e.g.,][]{Kokorev_2024a,Kocevski_2025}, the abundance of normal AGNs can be explained by a population of binary BHs at separations of $a\simeq 10^4~R_{\rm S}$ in our scenario.

Observationally, LRDs peak in abundance at $z\sim 6$ and decline sharply toward $z\lesssim 4$ \citep{Kocevski_2025,Ma_2025b}. 
This redshift evolution reflects the transient nature of LRDs with a characteristic fading timescale of $t_{\rm LRD} \lesssim 1~{\rm Gyr}$. 
\cite{Inayoshi_2025b} proposed that LRD phases correspond to the initial episode of AGN triggering processes and their unique features fade during the subsequent episodes as the system transitions into a typical AGN phase.
In our scenario, the activity of LRDs (more generally, AGNs) is driven by mergers of dark-matter halos and/or galaxies that host massive BHs.
Assuming a linear BH-to-halo mass relation, the BH mass ratio required to reproduce LRD spectra ($q\lesssim 0.1$; see Equation~\ref{eq:Luvopt} and discussion in Section~\ref{sec:3.1}) is naturally achieved through minor mergers of halos. 
Such mergers are frequent in cosmological $N$-body simulations, with merger rates per halo per unit redshift of order unity or higher for $q\lesssim 0.1$ \citep{Fakhouri_2010}, resulting in the emergence of LRDs in the early universe.

How do LRDs lose their characteristics in this framework?
One possible explanation is that their compact morphologies are disrupted by stellar mass build-up during subsequent mergers with normal galaxies. 
Star formation and stellar mass accumulation driven by these mergers could cause the AGN host to appear more extended.
Blending stellar components with the underlying AGN eases their unique LRD features.
\cite{Inayoshi_2025b} showed that a merger frequency of $p\gtrsim 0.5$ per halo per redshift interval (corresponding to mergers with $q\lesssim 0.1$)
can quantitatively explan the rapid declining trend of the LRD population \citep[see also][]{Kahn_2025}.

\subsection{So{\l}tan argument}\label{sec:soltan}

This scenario suggests that LRDs are not heavily obscured AGNs, but instead exhibit moderate extinction ($A_V\lesssim 1~{\rm mag}$).
Such weak reddening is further supported by the recent discovery of several LRDs showing a prominent Balmer break (e.g., \citealt{Naidu_2025,deGraaff_2025,Taylor_2025_z9.3}).
As a result, the intrinsic luminosity estimates for the entire LRD population are revised from 
$A_V \simeq 3-4$ to $A_V \lesssim 1$ mag.
This modification decreases the inferred intrinsic luminosity by a factor of $\sim 10~(\equiv 10^{\Delta A_V/2.5})$, 
where $\Delta A_V$ represents the difference of the attenuation level between our model and those assumed in
previous work that rely on heavy dust reddening
\citep[e.g.,][]{Kocevski_2023,Harikane_2023_agn,Matthee_2024,Greene_2024,Kokorev_2024b,Wang_2024b,Akins_2025}.
If this is the case, the luminosity density of LRDs is reduced by the same factor of 10, and the corresponding 
BH mass density decreases by a factor of $\sim 3$,
given that single-epoch virial mass estimates follow $M_{\rm BH} \propto L^{1/2}$, assuming the broad-line region luminosity-size empirical relation calibrated via reverberation mapping for nearby AGNs. 
Consequently, the radiative efficiency required to reconcile the BHAD and BH mass density for LRDs
is lowered by a factor of $\sim 3$ ($=10^{\Delta A_V/5.0}$), being consistent with the canonical 10\% radiative efficiency
\citep{Yu_Tremaine_2002,Ueda_2014,Delvecchio_2014} and thereby resolving the apparent tension
with the classical So{\l}tan argument \citep{Inayoshi_Ichikawa_2024}.

\subsection{Variability}\label{sec:var}

Flux variability is a key observational feature that supports the AGN origin of LRDs.
Based on limited multi-epoch observations (typically two or three visits per source), to date UV and optical data have revealed little or no significant variability among JWST-identified AGNs including LRDs \citep{Kokubo_Harikane_2025}.
An extended variability analysis using a larger sample of $\sim 300$ LRDs from publicly available photometric data 
also finds that most LRDs show no detectable variability on observational timescales, although eight sources are identified 
as variable LRD candidates with significant flux changes \citep{Zhang_2025}. 
Furthermore, some LRDs with spectroscopically confirmed broad Balmer emission lines exhibit variability on year-long 
timescales \citep{Ji_2025, Furtak_2025}.

In our model, flux variability arises from several components of the binary-disk system.
These include the orbital motion of the mini-disks and the interaction between shock-heated gas streams and both the outer edges 
of the mini-disks and the inner edge of the circum-binary disk \citep[e.g.,][]{Farris_2015,Munoz_2016,Miranda_2017,Tang_2018}.
Such interactions naturally produce multi-wavelength flux variations with strong periodicity, particularly at twice the binary orbital frequency with significant and distinctive chromaticity~\citep{Westernacher_2022}.
As a result, periodic variability on the timescale of a $\sim 2[(1+z)/5]M_7$ year in the observer frame emerges naturally when the binary separation is $a \sim 10^3~R_{\rm S}$ (see Equation~\ref{eq:fgw}), 
the regime where the characteristic LRD spectral features become prominent and persist over timescales of $\sim 0.1-1$ Myr (see Figure~\ref{fig:time}).
Confirming such periodicity will require multi-year baselines. However, Doppler modulations would produce periodicity in phase with any possible periodic red/blue-shifting of emission lines arising from the mini-disks (such as iron K$\alpha$ lines; \citealt{McKernan+2013}); spectroscopic time-domain observations could confirm these correlated continuum vs. line-shift modulations with a baseline covering less than a full period.

For simplicity, as mentioned above, we assume that the binary orbit remains circular throughout the evolution from the early binary-disk coupling phase to the final merger.
This assumption is justified because GW emission efficiently damps eccentricity during the late inspiral phase \citep{Peter_Mathews_1963}.
However, hydrodynamic simulations suggest that gas-driven interactions can excite and maintain orbital eccentricity \citep{Zrake_2021,Siwek_2023},
which potentially remain at moderate values of $e \sim \mathcal{O}(0.1)$ at the separations of $a \sim 10^3~R_{\rm S}$ relevant to the LRD formation.
Non-zero eccentricities can further modulate the light curves, introducing bursty signatures associated with accretion spikes near periapse \citep{Westernacher_2022}.
Future investigation is needed to examine the detailed connection between binary motion and the variability feature of LRDs 
in this framework.

\subsection{Multi-messenger observations for LRDs: implications to cosmology}

%%%%%%%
%   Fig.7   %
%%%%%%%
\begin{figure}
\centering
\includegraphics[width=84mm]{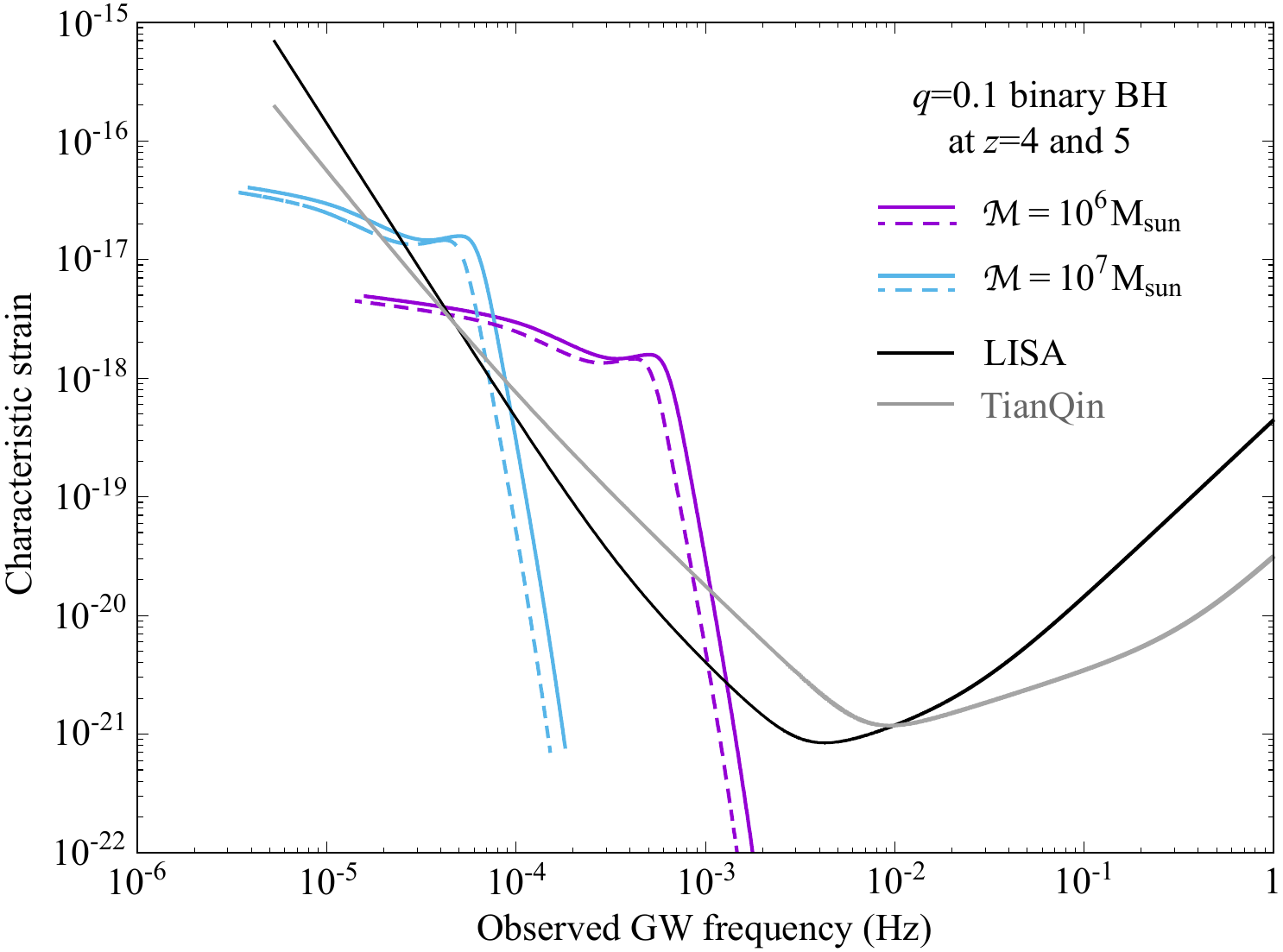}
\caption{GW spectra for non-spinning BH mergers with a mass ratio of $q=0.1$,
assuming two representative chirp masses $\mathcal{M}=10^6$ (purple) and $10^7~\msun$ (cyan), located at 
redshifts $z=4$ (solid) and $z=5$ (dashed), respectively. The sensitivity curves of LISA (black) and TianQin (gray) are overlaid.
}
\label{fig:gw}
\vspace{2mm}
\end{figure}

An intriguing question is whether binary BHs in the LRD phases can eventually be detectable as GW sources.
Figure~\ref{fig:gw} presents the GW spectra for non-spinning BH mergers with a mass ratio of $q=0.1$, assuming two representative chirp masses 
$\mathcal{M}=10^6$ and $10^7~\msun$, located at redshifts $z=4$ and $5$, respectively.
Each spectrum starts one year before merger, and is shown alongside the sensitivity curves of space-based GW detectors, 
LISA (black) and TianQin (gray) \citep{Amaro-Seoane_2023,TQ_2025}.

Binary BHs capable of producing LRD-like SEDs are expected to emit GW signals detectable by both LISA and TianQin during their final inspiral phases.
The higher-mass case ($\mathcal{M} = 10^7~\msun$) generates a GW signal that enters the detector 
frequency bands near the merger and ringdown phases, while the lower-mass system ($\mathcal{M} = 10^6~\msun$) 
may be observable over a longer duration from early inspiral to final coalescence.
The GW detectability improves with increasing mass ratio toward unity, as the signal shifts to higher frequencies. 
However, such nearly equal-mass binaries are unlikely to reproduce the distinct LRD SED, which appears to favor $q \sim 0.1$.
In contrast, systems with significantly smaller mass ratios ($q \ll 0.1$) may still preserve the LRD-like SED, depending on the accretion rate partition between 
the two BHs (i.e., $f_1$ and $f_2$) with $M\simeq 10^{6-8}~\msun$, but their GW signals shift below the low-frequency sensitivity limits of both detectors, making detection challenging.

As the binary orbit decays, the system eventually loses its distinct LRD features and enters a regime where it emits GWs detectable by LISA or TianQin.
However, this GW-emitting phase lasts much shorter than the LRD phase.
For our fiducial case of binary BHs with $M=10^7~\msun$ and $q=0.1$ at $z\simeq 4-5$, 
corresponding to a redshifted chirp mass of $\mathcal{M}_z= 2\times 10^6(1+z)~\msun$, 
only a small fraction, $\mathscr{F}\sim 10^{-6}(1+z)^{-8/3}$, of such binaries are expected to enter 
the LISA-detectable frequency band at $f_{\rm gw,min}\gtrsim 0.03~{\rm mHz}$ 
(see Figures~\ref{fig:time} and \ref{fig:gw}).
The binary coalescence time in the observer frame at this frequency is 
\begin{equation}
\tau_{\rm gw}\simeq 0.04~{\rm yr}~
\left(\frac{\mathcal{M}_z}{1.2\times 10^7~\msun}\right)^{-5/3}
\left(\frac{f_{\rm gw,min}}{0.03~{\rm mHz}}\right)^{-8/3}.
\end{equation}
Given the observed sky surface density of LRDs is $\Sigma_{\rm LRD}\sim 2,400~{\rm deg}^{-2}$ \citep{Kocevski_2025}, the expected event rate of high-$z$ GW-detectable binaries originating from LRDs over the full sky ($\mathcal{A}\simeq 40,000~{\rm deg}^2$) is estimated as $\mathscr{R}\simeq f_{\rm bin}\Sigma_{\rm LRD}\mathcal{A}\mathscr{F}/\tau_{\rm gw}$ or 
\begin{align}
\mathscr{R}
& \simeq 19 f_{\rm bin}~{\rm yr}^{-1} 
\left(\frac{\mathcal{M}}{2\times 10^6~\msun}\right)^{5/3}
\left(\frac{1+z}{6}\right)^{-1}
\\
&\times 
\left(\frac{f_{\rm gw,min}}{0.03~{\rm mHz}}\right)^{8/3}
\left(\frac{\Sigma_{\rm LRD}}{2,400~{\rm deg}^{-2}}\right)
\left(\frac{\mathcal{A}}{40,000~{\rm deg}^{2}}\right),\nonumber
\end{align}
assuming that all LRDs arise from binary BHs (i.e., $f_{\rm bin}=1$).
For a lower total BH mass with $M=10^6~\msun$, the fraction of $\mathscr{F}$ increases by a factor of $\sim 50$, which compensates for the $\mathcal{M}^{5/3}$ dependence in the coalescence time. As a result, the expected event rate remains comparable to $\sim 10 f_{\rm bin}~{\rm yr}^{-1}$.
This rate is broadly consistent with the theoretical predictions from previous studies in the literature based on cosmological galaxy simulations \citep{Kelley_2019,Volonteri_2020,Dong-Paez_2023}, merger-tree-based semi-analytical models \citep{Ricarte_2018,Barausse_2020,Liu_Inayoshi_2025}, and estimates calibrated by the abundance of low redshift quasars \citep{Xin_Haiman_2021}.

The Legacy Survey of Space and Time (LSST) by the Vera C. Rubin Observatory offers an ideal platform for identifying electromagnetic counterparts to GW sources originating from former LRD systems.
With more than five million exposures, LSST is collecting time-domain data across a deep, panoramic view of $\sim 18,000~{\rm deg}^2$ of the sky \citep{Ivezic_2019}\footnote{https://www.lsst.org/scientists/survey-design}, yielding the expected detection rate of such rare events of $\mathscr{R}\sim 9 f_{\rm bin}~{\rm yr}^{-1}$, depending on the binary fraction.
This estimate represents an upper limit, as the actual detection number is likely lower due to LSST's single-visit detection limit of $\sim 25$ mag in the $g$- and $r$-bands.
Therefore, this requires $\gtrsim 10$ visits to reach the brightness of typical faint AGNs identified by JWST \citep[e.g., CEERS~2782 at $z_{\rm spec}=5.242$ with $\sim 26$ mag;][]{Kocevski_2023}.
Consequently, the realistic detection rate of such binaries in joint observational programs is expected to be as low as $\sim \mathcal{O}(1)~{\rm yr}^{-1}$, which is broadly consistent with the predicted number of ultra-compact (orbital periods $\lesssim 1$ day) binary quasars computed from the luminosity function of low-redshift quasar populations \citep{Xin_Haiman_2021,Xin_Haiman_2024}.
A synergistic joint observation campaign with the Roman Space Telescope \citep{Spergel_2015}\footnote{https://science.nasa.gov/mission/roman-space-telescope/}, utilizing deeper multiple-band photometric follow-up for LSST-identified candidates, is expected to improve detectability.

\subsection{Stochastic GW background}

Another prediction of this scenario is the production of a stochastic gravitational wave background (GWB)
from the entire AGN population.
The characteristic GWB strain can be calculated as
\begin{equation}
h^2_{\rm c}=\frac{4}{3\pi^{1/3}}\frac{(G\mathcal{M})^{5/3}}{c^2 f_0^{4/3}}N_0 \langle (1+z)^{-1/3}\rangle,
\label{eq:hc}
\end{equation}
\citep{Phinney_2001}, where $f_0$ is the GW frequency in the observed frame, 
$N_0=\int_0^\infty N(z) dz$ is the present-day comoving number density of merged remnants, 
$N(z)$ is the event rate density per redshift bin between $z_{\rm min}$ and $z_{\rm max}$.
The averaged-redshift factor is defined as
\begin{equation}
\langle (1+z)^{-1/3}\rangle = \frac{1}{N_0}\int^{z_{\rm max}}_{z_{\rm min}}\frac{N(z)}{(1+z)^{1/3}}dz.
\end{equation}
We note that the formula in Equation~(\ref{eq:hc}) is valid under the assumption that the orbital evolution of the binary is dominated by GW emission rather than viscous torques.
The condition is typically satisfied for more massive binary BHs with tighter separations (see the binary-disk decoupling conditions in Figure~\ref{fig:time}).
Furthermore, if the binary orbit is eccentric, GW energy is emitted more efficiently at higher frequencies \citep[e.g.,][]{Enoki_2007,Sesan_2013,D'Orazio_2018},
though our analysis treats only circular orbits (see also discussion in Section~\ref{sec:var}).
The integration range is chosen to include the range $z_{\rm LRD} \simeq 4$–8, where the LRD occurrence rate peaks, 
yielding the GWB strength as
\begin{align}
h_{\rm c}\simeq &~5.6\times 10^{-17} \left(\frac{\mathcal{M}}{10^7~\msun}\right)^{5/6}
\left(\frac{f_0}{30~{\rm nHz}}\right)^{-2/3}\nonumber\\
&\times \left(\frac{N_0}{10^{-3}~{\rm Mpc}^{-3}}\right)^{1/2}
\left(\frac{\langle (1+z)^{-1/3}\rangle}{0.522}\right)^{1/2}.
\end{align}
Here, we assume that all LRDs originate from binary BHs with a certain fraction of the bulk AGN population,
including unobscured and obscured Type 2 AGNs with a total comoving number density of $N_0=10^{-3}~{\rm Mpc}^{-3}$.

The predicted GWB contribution from binary BHs powering LRDs accounts for $\sim 2\%$ of the observed strain amplitude 
$A_{\rm GWB}=2.4^{+0.7}_{-0.6}\times 10^{-15}$ (median with a 90\% credible interval) at a reference frequency 
of $1{\rm yr}^{-1}(=32~{\rm nHz})$, as reported by the NANOGrav 15-year dataset \citep{Agazie_2023_GWB,Agazie_2023_BBH}.
This is consistent with the recent conclusion in \citet{KisToth+2025}, which showed that if all AGN are associated with mergers, this yields approximately the observed GWB amplitude, dominated by AGN with BH masses of $\sim 10^9{\rm M_\odot}$ at redshifts $z\sim 2$.
The relatively low GWB amplitude from LRDs reflects the modest masses of binary BHs;
even at the high end, their chirp masses only reach $\mathcal{M} \sim 10^7~\msun$.
However, if this BH population undergoes further mass growth and experiences major mergers 
(e.g., reaching $\mathcal{M} \gtrsim 10^9~\msun$) in later epochs via gas-rich galaxy interactions, 
which are observed as (ultra-)luminous infrared galaxies \citep[e.g.,][]{Veilleux_2002,Delvecchio_2014}, 
the resulting GWB amplitude could approach the level detected by PTAs. 
This scenario predicts a GWB characteristic strain at $f_0=1~{\rm yr}^{-1}$ of 
$h_{\rm c}\simeq (1.24\pm 0.29) \times 10^{-15}$ \citep{Inayoshi_2018_GWB}, consistent with PTA measurements.

\section{Summary}\label{sec:5}

In this paper, we propose a scenario in which the v-shaped SED of LRDs originates from a binary BH system, 
where each BH is surrounded by a circum-BH mini-disk and embedded in a larger circum-binary disk. 
In this model, the mini-disks have higher effective temperatures, while the circum-binary disk remains cooler.
At a critical binary separation of $a\lesssim 10^3~R_{\rm S}$, the red optical emission originates from the Wien tail of 
a $T\simeq 5000~\K$ blackbody at the inner edge of the circum-binary disk, while the UV continuum is powered by the mini-disks. 
Binary torques carve out a gap between the circum-binary disk and mini-disks, setting the turnover wavelength of 
the v-shaped SED around the Balmer limit.

This configuration naturally reproduces LRD spectra without requiring strong dust attenuation ($A_V\lesssim 1$ mag). 
As an example, we apply the SED model to A2744-45924, the brightest known LRD at $z_{\rm spec} = 4.46$, which 
is well-constrained by JWST/NIRCam, MIRI, PRISM spectra, and ALMA upper limits 
\citep{Greene_2024,Labbe_2025,Setton_2025b}.
The model successfully fits its observed SED with only moderate extinction ($A_V = 0.6$ mag), 
significantly lower than previous estimates that required $A_V>2$ mag. 
While a more extensive application to the LRD population is left for future work, this paper instead focuses 
on establishing the theoretical framework.

The implication of moderate or negligible attenuation in LRDs provides a resolution to the overestimate
of AGN luminosities and thus alleviates the tension in the classical So{\l}tan argument. 
Previous analyses by \citet{Inayoshi_Ichikawa_2024} require radiative efficiencies of $\gtrsim 20-30\%$ 
to reconcile the inferred BHAD with the BH mass density derived by assuming significant dust attenuation
\citep{Matthee_2024,Greene_2024,Kokorev_2024b,Akins_2025}. 
In contrast, our model suggests that LRDs are consistent with a canonical 10\% efficiency, aligning with 
the value inferred from normal AGN populations at lower redshifts over $0<z<5$.

Finally, our scenario predicts spectral evolution driven by binary orbital decay due to interactions with 
the circum-binary disk and GW emission, linking early-stage ``proto-LRD" binaries to the broader AGN population 
and late-stage ``LRD-descendants" to coalescing binaries detectable in GW experiments;
space-based interferometers \citep{Amaro-Seoane_2023,TQ_2025,Liu_Inayoshi_2025} and PTA experiments 
\citep{Agazie_2023_GWB,Agazie_2023_BBH}.

\acknowledgments
We greatly thank Kenta Hotokezaka and Hanpu Liu for constructive discussions.
K.I., J.S., X.C., and L.C.H. acknowledge support from the National Natural Science Foundation of China (12573015, 1251101148, 12233001,12473037), 
the Beijing Natural Science Foundation (IS25003), and the China Manned Space Program (CMS-CSST-2025-A09).
J.S. is also supported by "The Fundamental Research Funds for the Central Universities, Peking University" (7100604896).
Z.H. acknowledges support by US NSF grant AST-2006176 and by NASA grants 80NSSC24K0440 and 80NSSC22K0822.
%\newpage

\appendix
\section{Disk Energetics}\label{sec:AppDisk}

Here, we describe the energy balance between the circum-binary disk and mini-disks,
and provide the conditions for reproducing the v-shaped SED in this binary-disk model.

In this model, the luminosity from the circum-binary disk is calculated as 
\begin{align}
L_{\rm cbd}&= -\int_{2a}^{\infty} \frac{3GM\dot{M}}{2r^2}dr = \frac{\dot{M}_0c^2}{20(1-p)}\left(\frac{a}{5R_{\rm S}}\right)^{p-1},\\
&\simeq 0.071~\dot{m}_0L_{\rm Edd}\left(\frac{a}{10^3~R_{\rm S}}\right)^{-1/2} \hspace{10pt}~{\rm for}~p=1/2.
\end{align}
The mini-disks are responsible for the UV emission.
The luminosity is estimated as 
\begin{align}
L_{{\rm md},i} &= -\int_{r_{{\rm in},i}}^{r_{{\rm out},i}} \frac{3GM_i\dot{M}_i}{2r^2}\left(1-\sqrt{\frac{r_{{\rm in},i}}{r}}\right)dr
= \frac{1}{12}f_i\dot{M}_0c^2,
\end{align}
where we assume $r_{{\rm in},i}\ll r_{{\rm out},i}$, which is valid for most cases we consider in this paper; $a\sim 10^3~R_{\rm S}$ and $q>0.01$.
Thus, the total mini-disk luminosity is $L_{{\rm md}}=L_{{\rm md},1}+L_{{\rm md},2}=(1/12)\dot{M}_0c^2$.
The luminosity density ratio between the far UV (from the primary mini-disk) and optical (from the circum-binary disk) is given for $p=1/2$ as
\begin{align}
\frac{L_{{\rm UV}}}{L_{{\rm opt}}}\simeq
\frac{L_{{\rm md,1}}}{L_{{\rm cbd}}}&=\frac{5(1-p)f_1}{3}\left(\frac{a}{5R_{\rm S}}\right)^{1-p},\\
&\simeq 1.2 \left(\frac{f_1}{0.1}\right)\left(\frac{a}{10^3~R_{\rm S}}\right)^{1/2},
 \hspace{10pt}~{\rm for}~p=1/2.
\label{eq:Lnuratio}
\end{align}
When keeping the turnover wavelength fixed at the Balmer limit, the ratio is rewritten as
\begin{equation}
\frac{L_{{\rm UV}}}{L_{{\rm opt}}}\simeq
1.03~\dot{m}_0^{1/5}M_7^{-1/5} \left(\frac{f_1}{0.1}\right)
\left(\frac{\lambda_{\rm t}}{3646~{\rm \AA}}\right)^{4/5}.
\end{equation}

\vspace{3mm}
\section{Imprinting a Balmer break on transmitted spectra}\label{sec:AppBalmer}

As discussed in \citet{Inayoshi_Maiolino_2025}, dense gas surrounding an AGN can imprint a Balmer break feature on the transmitted spectrum 
when the hydrogen number density of the gas slab is as high as $n_{\rm H} \simeq 10^{9-11}~\cc$, allowing atomic hydrogen to populate the $n = 2$ state.
We compute the attenuated AGN spectra with nebular emission for hydrogen column densities of 
$N_{\rm H} = 10^{23}$ and $10^{24}~{\rm cm}^{-2}$, assuming a hydrogen volume density of $n_{\rm H} = 10^{10}~{\rm cm}^{-3}$, an ionization parameter of $\log U = -1.5$,
and the incident AGN radiation spectrum as described in \citet{Inayoshi_Maiolino_2025}. 
We here consider a dust-free gas slab to isolate the effects of pure gas absorption and re-emission, and set the velocity of micro-turbulence to 
$v_{\rm tur}=100~\kms$ to make the Balmer break smooth \citep[e.g.,][]{Ji_2025}. 
The radiation transfer calculations are conducted with {\tt CLOUDY} \citep[C17,][]{Ferland_2017}.

To reproduce the overall continuum shape without resolving individual emission and absorption lines,
we provide a simple analytical function for the transmission curve with six parameters of $\lambda_0$, $\lambda_1$, and $s_i$ ($i=1,2,3,4$):
\begin{equation}
\mathcal{T}(\lambda)=
\left\{
\begin{array}{ll}
\exp \left[-\left(\dfrac{\lambda}{\lambda_0}\right)^{s_1} - \left(\dfrac{\lambda}{\lambda_1}\right)^{s_2}\right] & {\rm for}~\lambda<\lambda_{\rm B,lim},\\[7mm]
{s_3} \cdot \tanh \left[{s_4} \left(\dfrac{\lambda}{\lambda_{\rm B,lim}}-1\right)\right] & {\rm for}~\lambda \geq \lambda_{\rm B,lim},
\end{array}
\right.
\label{eq:BB_new}
\end{equation}
where $\lambda_0$ and $\lambda_1$ characterize the continuum attenuation due to gas absorption at wavelengths redward of 
the Ly$\alpha$ line and blueward of the Balmer limit ($\lambda_{\rm B,lim} = 3646~{\rm \AA}$), respectively.
The spectral shape redward of the Balmer limit is modeled using a hyperbolic tangent ($\tanh$) function with a parameter $s_4$ controlling 
the smoothness of the break.
This analytic expression offers a flexible description of the continuum shape modified by dense gas absorption, and is 
useful for fitting observed spectra where the Balmer break is a prominent feature.
For the two cases of $N_{\rm H} = 10^{23}$ and $10^{24}~{\rm cm}^{-2}$ with a covering fraction of 50\%, 
the best-fit parameters are
$(\lambda_0/{\rm \AA},\lambda_1/{\rm \AA},s_1,s_2,s_3,s_4)=(1293, 3965, -26.00, 2.337, 0.9905, 34.70)$ and $(1499, 2140,-12.44, 1.847, 0.8707, 26.64)$,
respectively.

\begin{figure*}
\centering
\includegraphics[width=134mm]{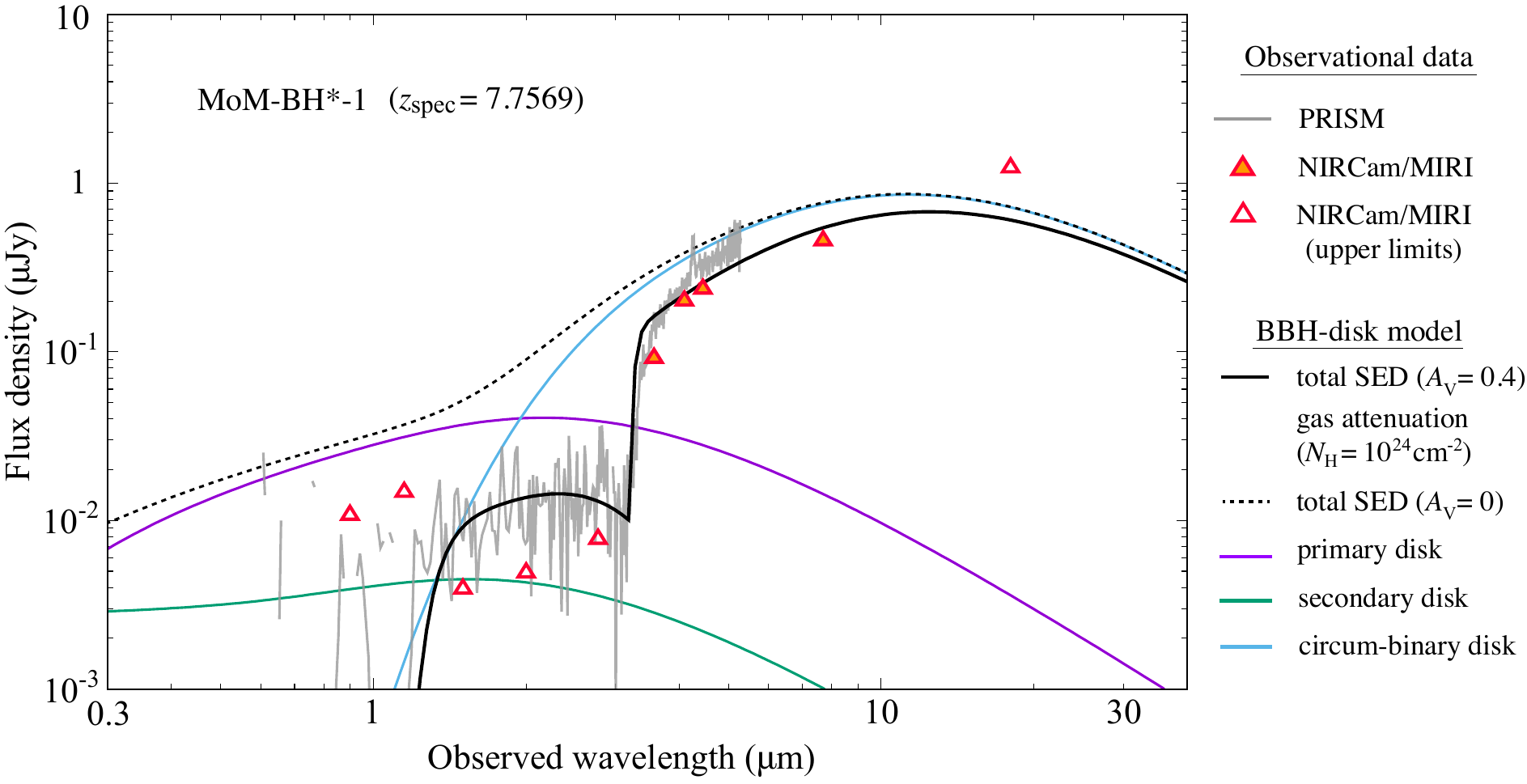}
\caption{A spectral model for the LRD with the deepest Balmer break, MoM-BH$^\star$-1, at $z_{\rm spec}=7.7569$
with NIRCam and MIRI photometry data and PRISM spectral data \citep{Naidu_2025}.
The binary BH-disk SED model assumes parameters $M=2\times 10^7~\msun$, 
$q=0.1$, $\dot{m}_0=3$, $p=0.75$, $a=10^3~R_{\rm S}$, and $A_V=0.4$.
The gas-attenuation transmission curve is adopted from the fitting formula in Equation~(\ref{eq:BB_new})
with $N_{\rm H}=10^{24}~{\rm cm}^{-2}$.}
\label{fig:MoM}
\vspace{2mm}
\end{figure*}

Figure~\ref{fig:MoM} shows the observational PRISM spectrum (gray line) and photometry data from NIRCam and MIRI (triangle)
of MoM-BH$^\star$-1 \citep{Naidu_2025}, which exhibits the strongest Balmer break among all spectroscopically observed LRDs.
The spectral model with the circum-binary disk and two mini-disks is computed for physical parameters of $M=2\times 10^7~\msun$, 
$q=0.1$, $\dot{m}_0=3$, $p=0.75$, $a=10^3~R_{\rm S}$, and $A_V=0.4$.
The gas-attenuation transmission curve is adopted from Equation~(\ref{eq:BB_new}) with $N_{\rm H}=10^{24}~{\rm cm}^{-2}$.

%%%%%%%
%   Fig.9   %
%%%%%%%

\bibliographystyle{aasjournal}
\bibliography{ref.bib}

%\bibliography{ref}{}
%\bibliographystyle{aasjournal}

\end{document}